\documentstyle[aps,eqsecnum]{revtex}
\begin{document}
\draft
\title{Radiative multipole moments of integer-spin fields in
       curved spacetime}
\author{Stephen W.~Leonard and Eric Poisson}
\address{Department of Physics, University of Guelph, Guelph,
         Ontario, N1G 2W1, Canada}
\date{Submitted to Physical Review D, May 8, 1997}
\maketitle

\begin{abstract}
Radiative multipole moments of scalar, electromagnetic, and linearized 
gravitational fields in Schwarzschild spacetime are computed to third 
order in $v$ in a weak-field, slow-motion approximation, where $v$ is
a characteristic velocity associated with the motion of the source. 
These moments are defined for all three types of radiation by relations
of the form $\Psi(t,\vec{x}) = r^{-1} \sum_{lm} {\cal M}_{lm}(u)\, 
Y_{lm}(\theta,\phi)$, where $\Psi$ is the radiation field at infinity
and ${\cal M}_{lm}$ are the radiative moments, functions of retarded
time $u=t-r-2M\ln(r/2M-1)$; $M$ is the mass parameter of the Schwarzschild
spacetime and $(t,\vec{x}) = (t,r,\theta,\phi)$ are the usual Schwarzschild 
coordinates. For all three types of radiation the moments share the same 
mathematical structure:

To zeroth order in $v$, the radiative moments are given by relations 
of the form ${\cal M}_{lm}(u) \propto (d/du)^l \int \rho(u,\vec{x})\, 
r^l\, \bar{Y}_{lm}(\theta,\phi)\, d\vec{x}$, where $\rho$ is the source 
of the radiation. A radiative moment of order $l$ is therefore given by
the corresponding source moment differentiated $l$ times with respect
to retarded time.

To second order in $v$, additional terms appear inside the spatial 
integrals, and the radiative moments become ${\cal M}_{lm}(u) \propto
(d/du)^l \int [1 + O(r^2 \partial_u^2) + O(M/r)]\, \rho\, r^l\, 
\bar{Y}_{lm}\, d\vec{x}$. The term involving $r^2\partial_u^2$ can be 
interpreted as a special-relativistic correction to the wave-generation 
problem. The term involving $M/r$ comes from general relativity. These 
correction terms of order $v^2$ are {\it near-zone} corrections which 
depend on the detailed behavior of the source. Furthermore, the radiative
multipole moments are still {\it local} functions of $u$, as they depend 
on the state of the source at retarded time $u$ only. 

To third order in $v$, the radiative moments become 
${\cal M}_{lm}(u) \to {\cal M}_{lm}(u) + 2M \int_{-\infty}^u 
[\ln(u-u') + \mbox{const}] \ddot{\cal M}_{lm}(u')\, du'$, where 
dots indicate differentiation with respect to $u'$. This
expression shows that the $O(v^3)$ correction terms occur outside the 
spatial integrals, so that they do not depend on the detailed behavior 
of the source. Furthermore, the radiative multipole moments now display 
a nonlocality in time, as they depend on the state of the source at
{\it all times} prior to the retarded time $u$, with the factor 
$\ln(u-u')$ assigning most of the weight to the source's recent past. 
(The term involving the constant is actually local.) The correction 
terms of order $v^3$ are {\it wave-propagation} corrections which are 
heuristically understood as arising from the scattering of the radiation 
by the spacetime curvature surrounding the source.

The radiative multipole moments are computed explicitly for all three 
types of radiation by taking advantage of the symmetries of the 
Schwarzschild metric to separate the variables in the wave equations. 
Our calculations show that the truly nonlocal wave-propagation 
correction --- the term involving $\ln(u-u')$ --- takes a universal 
form which is independent of multipole order and field type. We also 
show that in general relativity, temporal and spatial curvatures 
contribute {\it equally} to the wave-propagation corrections. Finally, 
we produce an alternative derivation of the radiative moments of a 
scalar field based on the retarded Green's function of DeWitt and Brehme.
This calculation shows that the tail part of the Green's function is
entirely responsible for the wave-propagation corrections in the radiative
moments. 
\end{abstract}

\section{Introduction and summary}

\subsection{Tails in waves}

It has long been known that in general, the propagation of massless 
fields in curved spacetime does not proceed along characteristics only, 
but is accompanied by wave tails. Much attention has been devoted to this 
topic since the ground-breaking work by Hadamard \cite{1}. 
Here are some of the highlights:

In 1952, Choquet-Bruhat \cite{2} studied the initial value problem of 
general relativity and showed that the gravitational field at 
some event P depends not only on the data put on the intersection 
of P's past light cone with the initial surface, but also on the 
data put inside this region. This result indicates that in general
relativity, field propagation proceeds at all speeds less than, or 
equal to, the speed of light. 

In 1960, DeWitt and Brehme \cite{3} constructed Green's functions for 
the scalar and electromagnetic wave equations in arbitrary curved 
spacetimes, and showed that these split naturally into a direct part, 
with support on, and only on, the light cone, and a tail part, with 
support within the light cone. A similar Green's function was also 
constructed for the Einstein field equations \cite{4}. 

In 1968, Kundt and Newman \cite{5} established that for hyperbolic 
partial differential equations in two dimensions, the presence of 
wave tails is the rule rather than the exception. This conclusion 
was extended by McLenaghan and co-workers \cite{6,6.1,6.2,6.3} 
to the case of conformally invariant wave equations in four dimensions. 

Wave tails are known to have important physical consequences.
For example, DeWitt and Brehme \cite{3} have shown that the tail 
part of the electromagnetic field is of paramount importance in 
deriving the equations of motion for charged particles in curved 
spacetime. Similarly, Mino, Sasaki, and Tanaka \cite{7}, as well 
as Quinn and Wald \cite{8}, have recently shown that tails are 
entirely responsible for the gravitational radiation reaction force. 
And as a final example, Price \cite{9} has shown that 
gravitational-wave tails play an integral part in the physical 
process by which a recently formed black hole relaxes to a 
stationary state, as is demanded by the no-hair theorems. 

The presence of tails in the gravitational waves produced by an 
isolated source was first demonstrated in 1965 by Bonnor and 
Rotenberg \cite{10}. In 1968, this work was extended by Couch, 
Torrence, Janis, and Newman \cite{11}, who showed that an initially 
outgoing wave will be partly backscattered by the spacetime curvature 
surrounding the source, thereby creating an incoming wave. A further 
extension of this work has appeared very recently \cite{12}. 

In 1992, Blanchet and Damour \cite{13} considered for the first time 
the effect of tails on the behavior of the gravitational waves at infinity, 
thereby concentrating on effects that could potentially be observed 
directly. They found that the gravitational waves at time $t$ depend 
not only on the state of the source at the corresponding retarded time 
$u$ [essentially $u=t-r/c$, where $r$ is the distance to the source, 
but see Eq.~(\ref{1.2}) for a more precise definition], but also on 
the state of the source at {\it all times} prior to the retarded time. 
(Once again, this indicates that wave propagation proceeds at all speeds 
less than, or equal to, the speed of light.) Subsequently 
\cite{14,15,16,17}, it was shown that tails play an important 
role in the generation of gravitational waves by the orbital 
inspiral of a compact binary system. These waves are among the most 
promising for detection by future kilometer-scale interferometers 
such as the American LIGO (Laser Interferometer Gravitational-wave 
Observatory \cite{18}) and the French-Italian VIRGO \cite{19}.

Given the physical relevance of tails in the propagation of radiation 
in curved spacetime, it appeared to us worthwhile to seek a deeper
understanding of this phenomenon by asking how it depends on the
type of radiation being considered, and by digging further into the 
nature of its physical origin. This paper reports on the results of 
such an investigation, in which we study the influence of tails 
on those properties of massless fields that are directly measurable 
to an observer at infinity: the radiative multipole moments. We 
consider the cases of scalar, electromagnetic, and gravitational 
radiation generated by an isolated source, and propagating to 
infinity in a spacetime curved by a nonrotating central mass $M$. 
All of our results are derived on the basis of a weak-field, 
slow-motion approximation. Throughout this paper we use units 
such that $G=c=1$, and we employ the definitions and conventions 
of Misner, Thorne, and Wheeler \cite{20}.

\subsection{Scalar radiation}

We begin the summary of our results with the simplest case, that
of a scalar field $\Phi(\bbox{x})$ obeying the wave equation 
$\Box \Phi = -4\pi \rho$, where $\Box = g^{\alpha\beta} 
\nabla_\alpha \nabla_\beta$, $g_{\alpha\beta}$ is the 
Schwarzschild metric, and $\rho(\bbox{x})$ a given source.
The symbol $\bbox{x}$ collectively designates all Schwarzschild
coordinates $\{t,r,\theta,\phi\}$. As is shown in Sec.~III, 
the radiative part of the scalar field, which dominates at 
infinity, can be written as
\begin{equation}
\Phi_{\rm rad}(t,\vec{x}) = 
\frac{1}{r} \sum_{l=0}^\infty \sum_{m=-l}^l 
{\cal Z}_{lm}(u) Y_{lm}(\theta,\phi),
\label{1.1}
\end{equation}
where ${\cal Z}_{lm}(u)$ are radiative multipole moments,
depending on retarded time
\begin{equation}
u = t - r - 2M \ln(r/2M - 1),
\label{1.2}
\end{equation}
and $Y_{lm}(\theta,\phi)$ are the usual spherical harmonics. The
symbol $\vec{x}$ collectively designates all spatial coordinates
$\{r,\theta,\phi\}$.

In a leading-order calculation in a weak-field, slow-motion
approximation, the radiative multipole moments are found to
be given by 
\begin{equation}
{\cal Z}^{(0)}_{lm}(u) = 
\frac{4\pi}{(2l+1)!!} \biggl( \frac{d}{du} \biggr)^l
\int \rho(u,\vec{x})\, r^l\, \bar{Y}_{lm}(\theta,\phi)\, d\vec{x},
\label{1.3}
\end{equation}
where $d\vec{x} = r^2\, dr\, d\cos\theta\, d\phi$ and the integration 
is over the region of space occupied by the source; we assume that 
this region is bounded. Equation (\ref{1.3}) shows that the radiative 
moments are obtained from the source moments 
$\int \rho\, r^l\, \bar{Y}_{lm}\, d\vec{x}$ by taking a number of
time derivatives equal to the multipole order.

In a more accurate calculation, incorporating corrections
of order $v^2$ with respect to the leading-order results (with
$v \ll 1$ a characteristic velocity associated with the motion 
of the source), we find
\begin{equation}
{\cal Z}^{(2)}_{lm}(u) = 
\frac{4\pi}{(2l+1)!!} \biggl( \frac{d}{du} \biggr)^l
\int \biggl[ 1 + \frac{(r\partial_u)^2}{2(2l+3)} - l\frac{M}{r} 
\biggr]\, \rho(u,\vec{x})\, r^l\, \bar{Y}_{lm}(\theta,\phi)\,
d\vec{x}.
\label{1.4}
\end{equation}
It is easy to see that the correction terms are indeed of order 
$v^2$. The term involving $(r\partial_u)^2$ is of order 
$(r_c/t_c)^2$, where $r_c$ is a characteristic radius within 
the source and $t_c$ a characteristic time scale associated 
with its motion; the ratio $r_c/t_c$ {\it defines} the characteristic
velocity $v$. This term can be understood as arising from 
special-relativistic corrections to the wave-generation problem.  
On the other hand, the term proportional to $M/r$ comes from general 
relativity, and is also of order $v^2$ by virtue of the virial 
theorem for bound motion in a gravitational field. It should be 
noted that in Eq.~(\ref{1.4}), the correction terms occur {\it inside}
the spatial integrals, so that they depend on the detailed
behavior of the source. Furthermore, these corrections are purely 
{\it local} in time, as ${\cal Z}^{(2)}_{lm}(u)$ depends on 
the state of the source at the time $u$ only. The terms of order 
$v^2$ are therefore {\it near-zone} corrections that have nothing 
to do with the tail effect discussed previously.

A calculation carried out to order $v^3$ in a weak-field,
slow-motion approximation does reveal the influence of the tails. 
Indeed, the radiative multipole moments are now given by
\begin{equation}
{\cal Z}^{(3)}_{lm}(u) = {\cal Z}^{(2)}_{lm}(u) + 
2M \int_{-\infty}^u \biggl[ \ln \biggl( \frac{u-u'}{4M} \biggr)  
+ \beta^{\rm scalar}_l + \gamma \biggr]\,
\ddot{\cal Z}^{(0)}_{lm}(u')\, du',
\label{1.5}
\end{equation}
which clearly displays a nonlocality in time. Here, a dot indicates 
differentiation with respect to $u'$, and 
\begin{equation}
\beta^{\rm scalar}_l = \psi(l+1) + \frac{1}{2},
\label{1.6}
\end{equation}
where $\psi(l+1) = -\gamma + \sum_{k=1}^l k^{-1}$ is the digamma
function ($\gamma \simeq 0.57721$ is Euler's number). We notice
that the correction terms of order $v^3$ occur {\it outside} the 
spatial integrals, so that they do not depend on the detailed behavior 
of the source. These are {\it wave-propagation} corrections, which
are readily associated with the occurrence of wave tails in curved 
spacetime. We must point out that of the three $O(v^3)$ terms in 
Eq.~(\ref{1.5}), two are actually {\it instantaneous}, as they are 
equal to $2M (\beta^{\rm scalar}_l + \gamma) 
\dot{\cal Z}^{(0)}_{lm}(u)$. The remaining term is truly nonlocal, 
and the factor $\ln(u-u')$ assigns most of the weight to the 
source's recent past. 

Equations (\ref{1.1})--(\ref{1.6}) were derived by integrating the 
wave equation for a scalar field in Schwarzschild spacetime, for 
which the variables can conveniently be separated. (The actual 
work of integrating the radial equation is carried out in Sec.~II.) 
But because our calculations use only the weak-field behavior of
the Schwarzschild solution, our results are insensitive to the 
detailed form of the metric. Although we rely heavily on the symmetries 
of the Schwarzschild metric to separate the variables in the wave 
equation, our results rely only on the fact that the field 
is spherically symmetric at large distances. Our results would 
therefore hold also in more general spacetimes, with spherical 
symmetry holding only approximately at large distances. Staticity, 
however, is a crucial assumption, and our results would not be
valid if the spacetime were rotating. Although the wave-propagation 
corrections to the radiative multipole moments would take 
the same form as in Eq.~(\ref{1.5}), the spacetime's rotation would 
create additional terms of order $v^3$. These would occur inside the 
spatial integrals, and would describe near-zone corrections of the 
spin-orbit type \cite{21,22,23}. 

\subsection{Electromagnetic radiation}

In Sec. IV we turn to the case of electromagnetic radiation 
produced by a given current density $J^\alpha(\bbox{x})$
in Schwarzschild spacetime. (The remarks of the preceding paragraph,
regarding the generality of our results, apply equally well here.) 
The radiative part of the vector potential is given by 
\begin{equation}
A^{\rm rad}_\alpha(t,\vec{x}) =  \frac{1}{r} 
\sum_{l=1}^\infty \sum_{m=-l}^l 
\Bigl[ {\cal I}_{lm}(u)\, Y^{E,lm}_\alpha (\theta,\phi) +
{\cal S}_{lm}(u)\, Y^{B,lm}_\alpha (\theta,\phi) \Bigr],
\label{1.7}
\end{equation}
where ${\cal I}_{lm}(u)$ and ${\cal S}_{lm}(u)$ are charge
and current multipole moments, respectively, while 
$Y^{E,lm}_\alpha (\theta,\phi)$ and 
$Y^{B,lm}_\alpha (\theta,\phi)$ are the vectorial spherical
harmonics described in Appendix A. In a calculation accurate
to order $v^3$ in a weak-field, slow-motion approximation, 
we find
\begin{equation}
{\cal I}^{(3)}_{lm}(u) = {\cal I}^{(2)}_{lm}(u) + 
2M \int_{-\infty}^u \biggl[\ln \biggl( \frac{u-u'}{4M} \biggr) 
+ \beta^{\rm em}_l + \gamma \biggr]\, 
\ddot{\cal I}^{(0)} _{lm}(u')\, du'
\label{1.8}
\end{equation}
and
\begin{equation}
{\cal S}^{(3)}_{lm}(u) = {\cal S}^{(2)}_{lm}(u) + 
2M \int_{-\infty}^u \biggl[\ln \biggl( \frac{u-u'}{4M} \biggr) 
+ \beta^{\rm em}_l + \gamma \biggr]\, 
\ddot{\cal S}^{(0)}_{lm}(u')\, du',
\label{1.9}
\end{equation}
where 
\begin{equation}
\beta^{\rm em}_l = \psi(l+1) + \frac{1}{2} - \frac{1}{2l(l+1)}.
\label{1.10}
\end{equation}
These relations are analogous to Eq.~(\ref{1.5}) and have the
same physical interpretation. The second-order multipole moments
are given by  
\begin{equation}
{\cal I}^{(2)}_{lm}(u) = 
\frac{4\pi}{(2l+1)!!}\, \sqrt{\frac{l+1}{l}} 
\biggl( \frac{d}{du} \biggr)^l \int 
\biggl[ 1 + \frac{l+3}{2(l+1)(2l+3)}\, (r \partial_u)^2 
 - 
(l-1) \frac{M}{r} \biggr] \sigma(u,\vec{x})\, 
r^l\, \bar{Y}_{lm}(\theta,\phi)\, d\vec{x}
\label{1.11}
\end{equation}
and
\begin{equation}
{\cal S}^{(2)}_{lm}(u) = -\frac{4\pi}{(2l+1)!!}\,
\biggl( \frac{d}{du} \biggr)^l 
\int \biggl[ 1 + \frac{(r \partial_u)^2}{2(2l+3)} - 
\frac{l^2-1}{l}\, \frac{M}{r} \biggr]\,
J^\alpha(u,\vec{x}) r^l\, 
\bar{Y}^{B,lm}_\alpha (\theta,\phi)\, d\vec{x},
\label{1.12}
\end{equation}
where $\sigma \equiv J^t - r \partial_u J^r/(l+1)$. 
These relations are analogous to Eq.~(\ref{1.4}) and 
have the same physical interpretation. Finally, the 
zeroth-order moments, ${\cal I}^{(0)}_{lm}(u)$
and ${\cal S}^{(0)}_{lm}(u)$, are obtained from Eqs.~(\ref{1.11})
and (\ref{1.12}) by discarding all $O(v^2)$ terms; in this
limit, $\sigma = J^{t}$. 

\subsection{Gravitational radiation}

The case of gravitational radiation is conceptually very different 
from the previous cases, because of the fact that the spacetime 
metric is now dynamical. However, if we assume that 
$T^{\alpha\beta}(\bbox{x})$, the stress-energy tensor responsible 
for the radiation, is small, then the Einstein field equations may 
be linearized in the small deviation of the metric with respect to 
the Schwarzschild form. This results in a wave equation for the 
metric perturbation \cite{24}, and mathematically, the 
gravitational-radiation problem ends up resembling closely the 
scalar and electromagnetic analogues. 
This is the problem considered in Sec.~V. 

The traceless-transverse gravitational-wave field is given by \cite{25}
\begin{equation}
h^{\rm rad}_{\alpha\beta}(t,\vec{x}) = \frac{1}{r} 
\sum_{l=2}^\infty \sum_{m=-l}^l
\Bigl[ {\cal I}_{lm}(u)\, T^{E2,lm}_{\alpha\beta}(\theta,\phi) 
+ {\cal S}_{lm}(u)\, T^{B2,lm}_{\alpha\beta}(\theta,\phi) \Bigr],
\label{1.13}
\end{equation}
where ${\cal I}_{lm}(u)$ and ${\cal S}_{lm}(u)$ are mass
and current multipole moments, respectively, while 
$T^{E2,lm}_{\alpha\beta}(\theta,\phi)$ and 
$T^{B2,lm}_{\alpha\beta}(\theta,\phi)$ are the tensorial spherical
harmonics described in Appendix A. The weak-field, slow-motion
approximations to the multipole moments are 
\begin{equation}
{\cal I}^{(3)}_{lm}(u) = {\cal I}^{(2)}_{lm}(u) + 
2M \int_{-\infty}^u \biggl[\ln \biggl( \frac{u-u'}{4M} \biggr) 
+ \beta^{\rm grav,mass}_l + \gamma \biggr]\, 
\ddot{\cal I}^{(0)}_{lm}(u')\, du'
\label{1.14}
\end{equation}
and
\begin{equation}
{\cal S}^{(3)}_{lm}(u) = {\cal S}^{(2)}_{lm}(u) + 
2M \int_{-\infty}^u \biggl[\ln \biggl( \frac{u-u'}{4M} \biggr) 
+ \beta^{\rm grav,current}_l + \gamma \biggr]\, 
\ddot{\cal S}^{(0)}_{lm}(u')\, du',
\label{1.15}
\end{equation}
where 
\begin{equation}
\beta^{\rm grav,current}_l = \psi(l+1) + \frac{1}{2} - \frac{2}{l(l+1)}
\label{1.16}
\end{equation}
and
\begin{equation}
\beta^{\rm grav,mass}_l = \beta^{\rm grav,current}_l - 
\frac{6}{(l-1)l(l+1)(l+2)}.
\label{1.17}
\end{equation}
Equations (\ref{1.14})--(\ref{1.16}) were previously derived (in a 
different representation involving symmetric tracefree tensors) by 
Blanchet \cite{26} (see also Refs.~\cite{13} and \cite{16}) on the
basis of post-Newtonian theory. Blanchet, however, derived these
relations without an explicit knowledge of the second-order moments,
which are given by 
\begin{equation}
{\cal I}^{(2)}_{lm}(u) = \frac{16\pi}{(2l+1)!!}\, 
\sqrt{ \frac{(l+1)(l+2)}{2(l-1)l} }\,
\biggl( \frac{d}{du} \biggr)^l
\int \biggl[
1 + \frac{l+9}{2(l+1)(2l+3)}\, (r \partial_u)^2 
- (l+2)\frac{M}{r} \biggr] \sigma(u,\vec{x})\, 
r^l\, \bar{Y}_{lm}(\theta,\phi)\, d\vec{x}
\label{1.18}
\end{equation}
and
\begin{eqnarray}
{\cal S}^{(2)}_{lm}(u) &=& - \frac{16\pi }{(2l+1)!!}\,
\sqrt{ \frac{2(l+2)}{l-1} }\, 
\biggl( \frac{d}{du} \biggr)^l
\int \biggl[ 
1 + \frac{l+4}{2(l+2)(2l+3)}\, (r \partial_u)^2 
\nonumber \\ & & \mbox{} 
- \frac{(l-1)(l+2)}{l}\, \frac{M}{r} \biggr]
J^\alpha(u,\vec{x})\, r^l\, 
\bar{Y}^{B,lm}_\alpha (\theta,\phi)\, d\vec{x},
\label{1.19}
\end{eqnarray}
where $\sigma \equiv T^{tt} + T^{rr} + r^2 T^{\theta\theta}
+ r^2\sin^2\theta T^{\phi\phi} - 4 r \partial_u T^{tr}/(l+1)$,
$J^\alpha \equiv T^{t\alpha} - r\partial_u T^{r\alpha}/(l+2)$.
To our knowledge, such explicit expressions for the radiative
multipole moments have not appeared before in the literature.
Finally, the zeroth-order expressions are recovered by discarding 
all $O(v^2)$ terms from Eqs.~(\ref{1.18}) and (\ref{1.19}); 
in this limit, $\sigma = T^{tt}$ and $J^\alpha = T^{t\alpha}$. 

The physical interpretation of these results is the same as 
in the previous cases, and the results share the same degree of 
generality as the previous ones (see the concluding paragraph 
of Sec.~I B). 

\subsection{Universality of the tail correction}

A survey of the preceding subsections reveals that the multipole 
moments of scalar, electromagnetic, and gravitational radiation fields 
all share the same mathematical structure, with terms of order $v^2$ 
near-zone corrections depending on the detailed behavior of the 
source, and with terms of order $v^3$ wave-propagation corrections 
independent of the detailed behavior of the source. And while the
near-zone corrections are local in time, the wave-propagation 
corrections are not. 

We also observe that the tail corrections depend on the multipole order 
$l$ and on the field's type only through the terms involving the 
various $\beta_l$'s. These terms are actually {\it instantaneous}, 
because after integration over $du'$, they are found to be 
proportional to the first derivative of the zeroth-order 
moments evaluated at $u$. The truly nonlocal tail corrections, which 
involve the weighting factor $\ln(u-u')$, are {\it independent} of 
multipole order and field type. This remarkable result, that the tail 
correction has a {\it universal} form, is one of the main new 
contributions of this paper. 

\subsection{Physical origin of the tail term}

The nonlocality (in time) of the radiative multipole moments is
heuristically understood as arising from the scattering of the
radiation field by the spacetime curvature surrounding the
mass $M$, and a survey of our previous results reveals that the
tail terms are indeed proportional to $M$. Now, the mass parameter 
enters twice in the metric of an asymptotically flat spacetime: 
Assuming that the weak-field metric is expressed in Schwarzschild-like 
coordinates, we have $g_{tt} \sim -1 +2M/r$ and $g_{rr} \sim 1 + 2M/r$ 
at large distances. Because of this degeneracy, it is impossible
to tell whether it is ``$g_{tt}$'s mass'' which is ``mostly
responsible'' for the tail effect, or whether it is ``$g_{rr}$'s
mass'', or whether both are ``equally responsible''. In other
words, we cannot tell how the temporal and spatial curvatures
separately contribute to the tail effect. 

We examine this question in Sec.~VI, for the specific case of
scalar radiation. To lift the degeneracy, we artificially introduce 
an additional mass parameter, $\hat{\gamma} M$, in the description 
of our spacetime. This is defined so that the metric functions are 
now given by $g_{tt} \sim -1 + 2M/r$ and $g_{rr} \sim 1 + 
2\hat{\gamma}M/r$ in the weak-field limit. General relativity is
recovered by setting $\hat{\gamma}=1$.

Integrating the scalar wave equation for the modified spacetime 
yields Eq.~(\ref{1.1}) with 
\begin{equation}
{\cal Z}^{(3)}_{lm}(u) = {\cal Z}^{(2)}_{lm}(u) + 
(1 + \hat{\gamma}) M \int_{-\infty}^u \biggl[ 
\ln \biggl( \frac{u-u'}{4M} \biggr)  
+ \beta^{\rm scalar}_l + \gamma \biggr]\,
\ddot{\cal Z}^{(0)}_{lm}(u')\, du',
\label{1.20}
\end{equation}
where $\beta^{\rm scalar}_l = \psi(l+1) + \hat{\gamma}/(1+
\hat{\gamma})$, 
\begin{equation}
{\cal Z}^{(2)}_{lm}(u) = 
\frac{4\pi}{(2l+1)!!} \biggl( \frac{d}{du} \biggr)^l
\int \biggl[ 1 + \frac{(r\partial_u)^2}{2(2l+3)} - 
\frac{(2l+1) \hat{\gamma} - 1}{2}\, \frac{M}{r} 
\biggr]\, \rho(u,\vec{x})\, r^l\, \bar{Y}_{lm}(\theta,\phi)\,
d\vec{x},
\label{1.21}
\end{equation}
and ${\cal Z}^{(0)}_{lm}(u)$ is given by Eq.~(\ref{1.3}). 
These expressions reduce to Eqs.~(\ref{1.5}) and (\ref{1.4}),
respectively, when $\hat{\gamma} = 1$. We see that in the 
modified spacetime, the tail terms are proportional to 
$(1+\hat{\gamma})$. This allows us to conclude that in 
general relativity, the temporal and spatial curvatures 
contribute {\it equally} to the tail effect. This intriguing 
observation is another main contribution of this paper. 

\subsection{Spacetime approach}

The final section of the paper, Sec.~VII, contains an alternative
derivation of Eqs.~(\ref{1.1})--(\ref{1.5}) based on the spacetime
approach of DeWitt and Brehme \cite{3}. For simplicity, we again 
restrict attention to the case of scalar radiation.

The mathematical methods employed in Secs.~III, IV, and V to derive
expressions for the radiative multipole moments of integer-spin fields 
are based upon a separation of variables approach made possible by
the symmetries of the Schwarzschild solution. These methods, though
convenient for practical computations, do not reflect closely the 
physical picture of wave propagation in curved spacetime. In particular,
the distinction between direct terms (which are local in time) and
tail terms (which are nonlocal) emerges only at the very end of the
calculation. 

The spacetime approach of Sec.~VII is based instead on 
$G(\bbox{x},\bbox{x}')$, the DeWitt-Brehme retarded Green's function 
for the scalar wave equation \cite{3}. As was mentioned in Sec.~I A, 
$G(\bbox{x},\bbox{x}')$ is expressed as a sum of two parts. 
The first part has support on, and only on, the past light cone 
of the field point $\bbox{x}$, and gives rise to the direct terms 
in the waves. The second part has support inside the past light 
cone of $\bbox{x}$, and gives rise to the tail terms. In the 
spacetime approach of DeWitt and Brehme, the mathematics reflect 
the physical picture quite closely.   

The radiative multipole moments calculated in Sec.~VII agree precisely 
with those calculated in Sec.~III. Therefore, the calculation based on
the spacetime approach tells us nothing new in terms of the final
answer. Nevertheless, this alternative derivation is very instructive, 
because of the fact that the mathematical origin of the tail correction 
is clear from the outset. We regard this as another important 
contribution of this paper.

\subsection{Organization of this paper}

The remaining sections of the paper contain the detailed derivations 
of the results summarized above. After laying some preliminary ground 
work in Sec.~II, we integrate the wave equations for scalar, 
electromagnetic, and gravitational radiation in Schwarzschild
spacetime in Secs.~III, IV, and V, respectively. All calculations
are carried out in a weak-field, slow-motion approximation. In 
Sec.~VI we integrate the scalar wave equation for the artificially 
modified spacetime. And finally, in Sec.~VII we integrate the 
scalar wave equation using the spacetime approach of DeWitt and 
Brehme \cite{3}. Various technical details are relegated to five 
Appendices.

\section{Generalized Regge-Wheeler equation}

The generalized Regge-Wheeler equation \cite{27},
\begin{equation}
\Biggl\{ \frac{d^2}{dr^{*2}} + \omega^2 - 
f \Biggl[ \frac{l(l+1)}{r^2} - \frac{2(s^2-1)M}{r^3}  \Biggl]
\Biggl\} X_l(\omega;r) = 0,
\label{2.1}
\end{equation}
has long been known to govern the evolution of 
integer-spin fields in Schwarzschild spacetime. Here, $r$ is 
the usual Schwarzschild coordinate, $f=1-2M/r$ (with $M$ 
denoting the mass of the spacetime), 
and $d/dr^* = f d/dr$. Also, $\omega$ 
denotes the frequency of the field, $l$ its multipole order, 
and $s = \{0,1,2\}$ its spin. The precise relation between the
mode functions $X_l(\omega,r)$ and the corresponding scalar, 
electromagnetic, and gravitational fields will be described
in Secs.~III, IV, and V, respectively. In this section we 
consider the purely mathematical problem of integrating 
Eq.~(\ref{2.1}) in the low-frequency limit.

We first examine the question of boundary conditions. It is 
easy to check that $X_l(\omega;r)$ must behave as $e^{\pm i \omega
r^*}$, where 
\begin{equation}
r^* = r + 2M \ln(r/2M - 1),
\label{2.3}
\end{equation}
in the asymptotic limits $r\to 2M$, $r\to\infty$. It will become 
clear in the following sections that the desired
solution is the one which describes purely incoming waves 
at the black-hole event horizon. We therefore select the 
function $X^H_l(\omega;r)$, such that 
\begin{equation}
X^H_l(\omega;r\to 2M) \sim ({\rm const})\, e^{-i \omega r^*}.
\label{2.4}
\end{equation}
The constant appearing in front of $e^{-i \omega r^*}$ determines
the overall normalization of the Regge-Wheeler function. Because
our final results will be independent of this normalization, we
shall leave this constant arbitrary. At infinity, $X^H_l(\omega;r)$ 
describes a superposition of incoming and outgoing waves. Consequently,
\begin{equation}
X^H_l(\omega;r\to\infty) \sim 
{\cal A}^{\rm in}_l(\omega) e^{-i\omega r^*}
+ {\cal A}^{\rm out}_l(\omega) e^{i\omega r^*}.
\label{2.5}
\end{equation}
The amplitudes ${\cal A}^{\rm in}_l(\omega)$ and 
${\cal A}^{\rm out}_l(\omega)$ are determined by 
solving the differential equation.

We wish to integrate Eq.~(\ref{2.1}) in the low-frequency
limit, for $M|\omega| \ll 1$. Without loss of generality,
we henceforth take $\omega$ to be positive; the
negative-frequency case can easily be recovered from
the relation $X^H_l(-\omega) = \bar{X}^H_l(\omega)$, where
a bar denotes complex conjugation. To facilitate the
calculations, we define the small (positive) quantity 
\begin{equation}
\varepsilon \equiv 2M\omega,
\label{2.6}
\end{equation}
and introduce a new dependent variable,
\begin{equation}
z = \omega r.
\label{2.7}
\end{equation}
After substitution and expansion in powers of $\varepsilon$,
the generalized Regge-Wheeler equation becomes
\begin{equation}
\Biggl\{ \frac{d^2}{dz^2} +  1 - \frac{l(l+1)}{z^2} 
+ \frac{\varepsilon}{z} \biggl[ \frac{1}{z} \frac{d}{dz} + 2
- \frac{l(l+1)+1-s^2}{z^2} \biggr] 
+ O(\varepsilon^2) \Biggr\} X_l(z) = 0.
\label{2.8}
\end{equation}
It should be noted that when expanding in powers of $\varepsilon$,
it was implicitly assumed that $z \gg \varepsilon$. Our low-frequency
solution will therefore be restricted to the domain $r \gg 2M$. 
Our task is now to integrate Eq.~(\ref{2.8}) to first order in
$\varepsilon$. We proceed by iteration, by writing $X_l = X_l^{(0)}
+ \varepsilon X_l^{(1)} + O(\varepsilon^2)$, substituting
into Eq.~(\ref{2.8}), and solving order by order. 

Because we are restricted to the domain $z \gg \varepsilon$, 
which excludes the event horizon at $z=\varepsilon$, Eq.~(\ref{2.4})
cannot be imposed directly, and the issue of boundary conditions must 
be re-examined. This question was addressed by Poisson
and Sasaki \cite{28}, who integrated the $s=2$ Regge-Wheeler 
equation in the domain $z \ll 1$ (which includes the horizon), 
imposed the correct boundary condition at $z=\varepsilon$, and then 
matched the resulting function to the general solution of Eq.~(\ref{2.8}) 
in the common domain $\varepsilon \ll z \ll 1$. Such an analysis will not 
be repeated here. It suffices to state the conclusion: To be
compatible with the incoming-wave boundary condition at the horizon,
the solution to Eq.~(\ref{2.8}) must be regular in the (unphysical, and
unrealized) limit $z \to 0$. 

With this in mind, the desired zeroth-order solution to 
Eq.~(\ref{2.8}) is
\begin{equation}
X^{H(0)}_l(z) = z j_l(z),
\label{2.9}
\end{equation}
where $j_l(z)$ are the spherical Bessel functions of the first kind. 
It should be noted that Eq.~(\ref{2.9}) provides a particular
choice for the overall normalization of $X^H_l(z)$. The
first-order solution is then determined by solving
\begin{equation}
\biggl[ \frac{d^2}{dz^2} +  1 - \frac{l(l+1)}{z^2} \biggr] X^{(1)}_l(z)
= -W_l(z),
\label{2.10}
\end{equation}
where
\begin{equation}
W_l(z) = \frac{1}{z} \biggl[ \frac{1}{z} \frac{d}{dz} + 2
- \frac{l(l+1)+1-s^2}{z^2} \biggr] z j_l(z).
\label{2.11}
\end{equation}
The general solution to Eq.~(\ref{2.10}) is
\begin{equation}
X^{(1)}_l(z) = z j_l(z) 
\biggl[ a + \int^z z' n_l(z') W_l(z')\, dz' \biggr]
+ z n_l(z) \biggr[ b - \int^z z' j_l(z') W_l(z')\, dz' \biggr],
\label{2.12}
\end{equation}
where $n_l(z)$ are the spherical Bessel functions of the second kind,
and $a$ and $b$ are constants which must be chosen so that
$X^{(1)}_l(z)$ satisfies the regularity condition at $z=0$.

The integrations of Eq.~(\ref{2.12}) can be carried out explicitly.
First, we use the recurrence relations among spherical Bessel 
functions (Ref.~\cite{29}, p.~439) to write $W_l(z)$ in the form
\begin{equation}
z W_l(z) = 2 z j_l(z) - \frac{(l-s)(l+s)}{2l+1}\, j_{l-1}(z) 
 - \frac{(l-s+1)(l+s+1)}{2l+1}\, j_{l+1}(z).
\label{2.13}
\end{equation}
Second, we evaluate the integrals using the results found in the 
Appendix of Ref.~\cite{30}. After straightforward manipulations, we
arrive at
\begin{eqnarray}
X^{(1)}_l(z) &=& \Bigl[ a - A_l(z) \Bigr] z j_l(z) + 
\Bigl[ b + \gamma + B_l(z) \Bigr] z n_l(z) 
 - \frac{(l-s)(l+s)}{2l(2l+1)}\, z j_{l-1}(z) 
\nonumber \\ & & \mbox{} +
\frac{(l-s+1)(l+s+1)}{2(l+1)(2l+1)}\, z j_{l+1}(z),
\label{2.14}
\end{eqnarray}
where
\begin{equation}
A_l(z) = {\rm Si}(2z) + z^2 n_0(z) j_0(z) 
+ \sum_{p=1}^{l-1}
\biggl(\frac{1}{p} + \frac{1}{p+1} \biggr) z^2 n_p(z) j_p(z)
\label{2.15}
\end{equation}
and
\begin{equation}
B_l(z) = {\rm Ci}(2z) - \gamma - \ln(2z) + z^2 {j_0}^2(z) 
+ \sum_{p=1}^{l-1}
\biggl(\frac{1}{p} + \frac{1}{p+1} \biggr) z^2 {j_p}^2(z).
\label{2.16}
\end{equation}
Here, Si and Ci are the sine and cosine integral functions, 
respectively, and $\gamma \simeq 0.57721$ is Euler's number. 

In Appendix B, the functions $A_l(z)$ and $B_l(z)$ are evaluated 
in the limit $z \to 0$. We find
\begin{equation}
A_l(z) = \frac{z}{l} + O(z^3), \qquad
B_l(z) = -\frac{z^{2l+2}}{l(l+1)(2l-1)!!(2l+1)!!} + O(z^{2l+4}).
\label{2.17}
\end{equation}
It follows that $X^{(1)}(z)$ goes to zero in the limit $z\to 0$
provided that $b = -\gamma$. Otherwise, the function diverges. 
This choice for $b$ therefore selects $X^{H(1)}_l(z)$, the desired 
solution. The constant $a$ remains arbitrary, because it affects 
only the overall normalization of the solution; $a$ can be set to 
zero without loss of generality.

Integration of the generalized Regge-Wheeler function, to first order
in $\varepsilon$, is now completed. We have pointed out that our answer 
incorporates a specific choice of overall normalization which is 
provided by Eq.~(\ref{2.9}). We now wish to form a 
normalization-independent quantity, 
$X^H_l(z)/{\cal A}^{\rm in}_l$, which will be 
required in the following sections of this paper. We must therefore 
calculate ${\cal A}^{\rm in}_l$. This 
involves the evaluation of $X^{H}_l (z)$, as given by 
Eqs.~(\ref{2.9}) and (\ref{2.14})--(\ref{2.16}), in the
limit $z \to \infty$, and a comparison with the low-frequency
limit of Eq.~(\ref{2.5}). Such a calculation is carried out in 
Appendix C. The final result is quoted here:
\begin{eqnarray}
\frac{X^H_l(z)}{{\cal A}^{\rm in}_l} &=& 2 (-i)^{l+1} 
e^{i\varepsilon (\ln 2\varepsilon - \beta_l)} 
\bigl(1+ {\textstyle \frac{\pi}{2}} \varepsilon\bigr) z 
\Biggl\{ \Bigl[ 1 - \varepsilon A_l(z) \Bigr] j_l(z) + 
\varepsilon B_l(z) n_l(z) 
\nonumber \\ & & \mbox{} - 
\varepsilon \frac{(l-s)(l+s)}{2l(2l+1)}\, j_{l-1}(z) +
\varepsilon \frac{(l-s+1)(l+s+1)}{2(l+1)(2l+1)}\, j_{l+1}(z)
+ O(\varepsilon^2) \Biggr\},
\label{2.18}
\end{eqnarray}
where 
\begin{equation}
\beta_l = \psi(l+1) + \frac{1}{2} - \frac{s^2}{2l(l+1)},
\label{2.19}
\end{equation}
with $\psi(l+1) = -\gamma + \sum_{k=1}^l k^{-1}$ denoting the 
digamma function.

Finally, we evaluate Eq.~(\ref{2.18}) in the limit $z \ll 1$, to
a degree of accuracy sufficient for our purposes in the following
sections of this paper. For this, we use Eq.~(\ref{2.17}) and the 
series expansions for the spherical Bessel functions. We arrive at
\begin{eqnarray}
\frac{X^H_l(\omega;r)}{{\cal A}^{\rm in}_l(\omega)} &=& 
\frac{2}{(2l+1)!!}\, 
e^{2 i M \omega (\ln 4 M |\omega| - \beta_l)} 
\bigl(1+ \pi M |\omega|\bigr) (-i\omega r)^{l+1} 
\nonumber \\ & & \mbox{} \times
\Biggl\{
1 - \frac{(\omega r)^2}{2(2l+3)} - \frac{(l-s)(l+s)}{l}\, \frac{M}{r}
+ O\Bigl[ (\omega r)^4, M\omega^2 r, (M/r)^2\Bigr] 
\Biggr\}.
\label{2.20}
\end{eqnarray}
This equation holds both for positive and negative frequencies.

\section{Scalar radiation}

\subsection{Wave equation}

We begin our study of radiative multipole moments in curved spacetime 
with the simplest case, that of a real scalar field $\Phi(\bbox{x})$ 
obeying the wave equation
\begin{equation}
\Box \Phi (\bbox{x}) = -4\pi \rho(\bbox{x}).
\label{3.1}
\end{equation}
Here, $\Box = g^{\alpha\beta} \nabla_{\alpha} \nabla_{\beta}$ is
the curved spacetime wave operator, $\rho(\bbox{x})$ is an unspecified 
source function, and $\bbox{x}$ collectively designates all spacetime 
coordinates. The spacetime is assumed to be Schwarzschild (with mass $M$), 
and the usual coordinates $\{t,r,\theta,\phi\}$ are adopted. 

Because the spacetime is static and spherically symmetric, the
scalar field can be decomposed according to
\begin{equation}
\Phi(\bbox{x}) = \frac{1}{r} \int d\omega \sum_{lm} R_{lm}(\omega;r)\,
Y_{lm}(\theta,\phi) e^{-i\omega t},
\label{3.2}
\end{equation}
where the sums over $l$ and $m$ are restricted by $l\geq 0$, 
$|m| \leq l$. Substituting this into Eq.~(\ref{3.1}), we obtain 
the following ordinary differential equation for the radial function 
$R_{lm}(\omega;r)$:
\begin{equation}
\Biggl\{ \frac{d^2}{dr^{*2}} + \omega^2 - f \biggl[ \frac{l(l+1)}{r^2} 
+ \frac{2M}{r^3} \biggr] \Biggr\} R_{lm}(\omega;r)
= f T_{lm}(\omega;r),
\label{3.3}
\end{equation}
where $d/dr^* = f d/dr$ and $f=1-2M/r$. The source term is given by
\begin{equation}
T_{lm}(\omega;r) = -4\pi r \int \tilde{\rho}(\omega,\vec{x})\, 
\bar{Y}_{lm}(\theta,\phi)\, d\Omega,
\label{3.5}
\end{equation}
where $d\Omega = d\cos\theta\, d\phi$, a bar denotes complex 
conjugation, and
\begin{equation}
\tilde{\rho}(\omega,\vec{x}) = \frac{1}{2\pi} \int \rho(t,\vec{x})\,
e^{i\omega t}\, dt
\label{3.6}
\end{equation}
is the Fourier transform of $\rho(\bbox{x})$. The symbol
$\vec{x}$ collectively designates all spatial coordinates. 

\subsection{Solution}

Equation (\ref{3.3}) has a Sturm-Liouville form, and it can therefore
be solved in terms of a Green's function constructed from two linearly
independent solutions to the homogeneous problem. Which solutions 
are selected depends on the boundary conditions we wish to impose 
on $R_{lm}(\omega;r)$. The appropriate choice here is dictated by the
physical requirement that the scalar field must represent waves which 
are purely ingoing at the black hole horizon ($r=2M$), and purely 
outgoing at $r=\infty$. This amounts to integrating Eq.~(\ref{3.1}) 
with a no-incoming-radiation initial condition.

We therefore seek functions 
$R^H_{l}(\omega;r)$ and $R^\infty_{l}(\omega;r)$, solutions to
\begin{equation}
\Biggl\{ \frac{d^2}{dr^{*2}} + \omega^2 - f \biggl[ \frac{l(l+1)}{r^2} 
+ \frac{2M}{r^3} \biggr] \Biggr\}  R_{l}(\omega;r) = 0,
\label{3.7}
\end{equation}
and such that
\begin{eqnarray}
R^H_{l}(\omega; r\to 2M) &\sim& e^{-i\omega r^*}, \nonumber \\
R^H_{l}(\omega; r\to\infty) &\sim& 
{\cal Q}^{\rm in}_l(\omega) e^{-i\omega r^*} + 
{\cal Q}^{\rm out}_l(\omega) e^{i\omega r^*}, \label{3.8} \\
R^\infty_{l}(\omega; r\to\infty) &\sim& e^{i\omega r^*}, \nonumber
\end{eqnarray}
where $r^* = r + 2M\ln(r/2M - 1)$. Equation (\ref{3.8}) indicates 
that $R^H_l(\omega;r)$ describes waves which are purely ingoing
at the black-hole horizon, while $R^\infty_l(\omega;r)$ describes
waves which are purely outgoing at infinity. The behavior of 
$R^\infty_l(\omega;r)$ near $r=2M$ will not be needed.  

In terms of these functions, the solution to Eq.~(\ref{3.3}) 
takes the form
\begin{eqnarray}
R_{lm}(\omega; r) &=& \frac{1}{2i\omega {\cal Q}^{\rm in}_l(\omega)}\,
\Biggl[ R^\infty_l(\omega;r) 
\int_{2M}^{r} T_{lm}(\omega;r') R^H_l(\omega;r')\, dr' 
\nonumber \\ & & \mbox{} 
+ R^H_l(\omega;r)
\int_{r}^{\infty} T_{lm}(\omega;r') R^\infty_l(\omega;r')\, dr' \Biggr],
\label{3.9}
\end{eqnarray}
where the factor $2i\omega {\cal Q}^{\rm in}_l(\omega)$ is the 
conserved Wronskian of the functions $R^H_{l}(\omega;r)$ and 
$R^\infty_{l}(\omega;r)$.

We are interested in the radiative part of the field, which dominates
at large distances from the source. More precisely, we define
the radiative field by $r\Phi_{\rm rad}(\bbox{x}) = \lim_{r\to\infty}
r \Phi(\bbox{x})$, which expresses the fact that at large distances,
$\Phi(\bbox{x}) = \Phi_{\rm rad}(\bbox{x}) + O(1/r^2)$. Evaluating 
Eq.~(\ref{3.9}) in the limit $r\to\infty$ [assuming that $T_{lm}(\omega;r)$ 
has compact support], we obtain 
\begin{equation}
R_{lm}(\omega;r\to\infty) \sim \tilde{\cal Z}_{lm}(\omega)\, 
e^{i\omega r^*},
\label{3.10}
\end{equation}
where
\begin{equation}
\tilde{\cal Z}_{lm}(\omega) \equiv 
\frac{1}{2i\omega {\cal Q}^{\rm in}_l(\omega)}\,
\int_{2M}^{\infty} T_{lm}(\omega;r) R^H_l(\omega;r)\, dr.
\label{3.11}
\end{equation}
Finally, substituting this into Eq.~(\ref{3.2}) yields
\begin{equation}
\Phi_{\rm rad}(t,\vec{x}) = 
\frac{1}{r} \sum_{lm} {\cal Z}_{lm}(u) Y_{lm}(\theta,\phi),
\label{3.12}
\end{equation}
where $u=t-r^*$ is retarded time, and
\begin{equation}
{\cal Z}_{lm}(u) = \int \tilde{\cal Z}_{lm}(\omega) 
e^{-i\omega u}\, d\omega.
\label{3.13}
\end{equation}
The quantities ${\cal Z}_{lm}(u)$, or their Fourier transform 
$\tilde{\cal Z}_{lm}(\omega)$, will be referred to as the {\it 
radiative multipole moments} of the scalar field $\Phi(\bbox{x})$. 

\subsection{Slow-motion approximation}

In Eq.~(\ref{3.11}), the radiative multipole moments are written
in exact form in terms of the source $T_{lm}(\omega;r)$ and the
function $R^H_l(\omega;r)/{\cal Q}^{\rm in}_l(\omega)$. While
the source function will be left unspecified, we now wish to 
find an expression for 
$R^H_l(\omega;r)/{\cal Q}^{\rm in}_l(\omega)$. To do this
we must resort to approximations, because Eq.~(\ref{3.7}) 
cannot be integrated in closed form. 

We will derive approximate expressions for the radiative multipole 
moments, and these will be valid in weak-field, slow-motion situations. 
To formulate this approximation precisely, we introduce a characteristic 
radius $r_c$, to be thought of as the radial coordinate of a typical 
portion of the source. [In other words, $T_{lm}(\omega;r)$ is assumed
to be appreciably different from zero only for values of $r$ comparable
to $r_c$.] We introduce also a characteristic time $1/\omega_c$, to be 
thought of as the typical time scale over which the source moves. [In
other words, $T_{lm}(\omega;r)$ is assumed to be appreciably different
from zero only for values of $\omega$ comparable to $\omega_c$.] Finally, 
we introduce a characteristic velocity $v \ll 1$, which will be the  
smallness parameter of our approximation. In terms of these quantities, 
the requirement that the source motions must be slow translates to 
\begin{equation}
\omega_c r_c = O(v). 
\label{3.14}
\end{equation}
The virial theorem for gravitationally bound systems then implies 
that the gravitational field must be weak inside the source:
\begin{equation}
M/r_c = O(v^2). 
\label{3.15}
\end{equation}
Finally, the slow-motion approximation implies that the scalar waves 
produced by the source's motion must have low frequencies: 
\begin{equation}
M \omega_c = O(v^3). 
\label{3.16}
\end{equation}
Our calculation of the radiative multipole moments will be carried out 
to order $v^3$ beyond the leading-order expressions.

We now proceed. It is evident that Eq.~(\ref{3.7}) is nothing but 
Eq.~(\ref{2.1}), the generalized Regge-Wheeler equation, with $s=0$. 
So we have, immediately,
\begin{equation}
\frac{R^H_l(\omega;r)}{{\cal Q}^{\rm in}_l(\omega)} = 
\frac{X^H_l(\omega;r)}{{\cal A}^{\rm in}_l(\omega)}.
\label{3.17}
\end{equation}
Equation (\ref{2.20}) may therefore be substituted into Eq.~(\ref{3.11}).
After using Eq.~(\ref{3.5}), we obtain
\begin{eqnarray}
\tilde{\cal Z}_{lm}(\omega) &=& \frac{4\pi}{(2l+1)!!} 
e^{2 i M \omega (\ln 4 M |\omega| - \beta_l)} 
\bigl(1+ \pi M |\omega|\bigr) (-i\omega)^l 
\nonumber \\ & & \mbox{} \times
\int \biggl[ 1 -
\frac{(\omega r)^2}{2(2l+3)} - l\frac{M}{r} 
+ O(v^4) \biggr]\, \tilde{\rho}(\omega,\vec{x})\,
r^l\, \bar{Y}_{lm}(\theta,\phi)\, d\vec{x},
\label{3.18}
\end{eqnarray}
where 
\begin{equation}
\beta_l = \psi(l+1) + \frac{1}{2},
\label{3.19}
\end{equation}
and $d\vec{x} = r^2 dr d\Omega$.

The physical interpretation of this result comes more easily if
we first invert the Fourier transform. The $\omega$-dependent prefactors
complicate this procedure slightly, but an explicit expression for
${\cal Z}_{lm}(u)$ can nevertheless be found (see Appendix D). 
We obtain
\begin{equation}
{\cal Z}_{lm}(u) = Z_{lm}(u) + 2M \int_{-\infty}^u 
\biggl[ \ln \biggl( \frac{u-u'}{4M} \biggr)  
+ \beta_l + \gamma \biggr]\,
\ddot{Z}_{lm}(u')\, du',
\label{3.20}
\end{equation}
where a dot indicates differentiation with respect to $u'$, and
\begin{equation}
Z_{lm}(u) = \frac{4\pi}{(2l+1)!!} \biggl( \frac{d}{du} \biggr)^l
\int \biggl[ 1 + \frac{(r\partial_u)^2}{2(2l+3)} - l\frac{M}{r} 
+ O(v^4) \biggr]\, \rho(u,\vec{x})\, r^l\, \bar{Y}_{lm}(\theta,\phi)\,
d\vec{x}.
\label{3.21}
\end{equation}
Equation (\ref{3.21}) indicates that to leading order, 
${\cal Z}_{lm}(u)$ is given by the $l$-th retarded-time derivative of 
$\int \rho r^l \bar{Y}_{lm}\, d\vec{x}$. This justifies our 
referring to these quantities as multipole moments. Equation (\ref{3.21})
also shows that all corrections of order $v^2$ appear {\it inside} the
spatial integral, and that they depend on the detailed behavior of
the source function. These corrections are {\it near-zone} corrections,
and they are purely local in time: as with the leading-order term, they 
involve the value of the source at the retarded time $u$ only. This is 
not so for the corrections of order $v^3$, as is shown by 
Eq.~(\ref{3.20}): these involve the value of the source at {\it all times} 
prior to the retarded time $u$. Furthermore, the $O(v^3)$ corrections appear 
{\it outside} the spatial integral, and they are independent of the
detailed behavior of the source. These are {\it wave-propagation} 
corrections. Equations (\ref{3.20}) and (\ref{3.21}), with $\beta_l$
given by Eq.~(\ref{3.19}), are equivalent to Eqs.~(\ref{1.4})--(\ref{1.6}).

We recognize the important distinction between near-zone and 
wave-propagation corrections. Near-zone corrections depend
on the detailed behavior of the source and are local in
time. Wave-propagation corrections, on the other hand, do
not depend on the detailed behavior of the source, and are
nonlocal in time. This nonlocality is heuristically understood
as arising from the scattering of the radiation by the spacetime 
curvature surrounding the source. This scattering causes part of
the information about the state of the source to be delayed further 
than what is strictly required by causality. The integral term in 
Eq.~(\ref{3.20}) is often called the {\it tail} term, and 
wave-propagation corrections are often called {\it tail} corrections. 

\section{Electromagnetic radiation}

\subsection{Teukolsky equation}

In this section, we derive expressions for the radiative multipole 
moments of an electromagnetic field in Schwarzschild spacetime. 
The Maxwell equations for this spacetime 
have been cast, by Teukolsky \cite{31}, in a form convenient for 
our purposes. In the Teukolsky formalism, the radiative part of the 
electromagnetic field is represented by the complex quantity $\Phi_2 
= F^{\alpha\beta} \bar{m}_\alpha n_\beta$, 
where $F^{\alpha\beta}$ is the field tensor, 
$n_\alpha = -\frac{1}{2}(f,1,0,0)$ a null vector pointing 
radially inward, and 
$\bar{m}_\alpha = (0,0,r,-ir\sin\theta)/\sqrt{2}$ 
a spatial vector with zero norm. As before, a bar
denotes complex conjugation.

The field $\Phi_2(\bbox{x})$ has spin-weight $s=-1$ (see Appendix A),
and it can be decomposed according to 
\begin{equation}
\Phi_2(\bbox{x}) = \frac{1}{r^2} \int d\omega \sum_{lm} 
R_{lm}(\omega;r)\, \mbox{}_{-1} Y_{lm}(\theta,\phi)\, 
e^{-i\omega t},
\label{4.1}
\end{equation}
where $\mbox{}_{-1} Y_{lm}(\theta,\phi)$ are spherical harmonics
of spin-weight $-1$ (see Appendix A). The sums over
$l$ and $m$ are restricted by $l \geq 1$, $|m| \leq l$. The radial
function then satisfies the inhomogeneous Teukolsky 
equation \cite{31},
\begin{equation}
\Biggl\{ r^2 f \frac{d^2}{dr^2} + \frac{1}{f} \Bigl[
(\omega r)^2 - 2i\omega r (1-3M/r) \Bigr] - l(l+1) \Biggr\} 
R_{lm}(\omega;r) = T_{lm}(\omega;r),
\label{4.2}
\end{equation}
where $f=1-2M/r$. 

The source term, $T_{lm}(\omega;r)$, is constructed as follows 
from $J^\alpha(\bbox{x})$, the (unspecified) current 
density. One first forms the contractions 
\begin{equation}
\mbox{}_{0} J = - J^\alpha n_\alpha, \qquad
\mbox{}_{-1} J = - J^\alpha \bar{m}_\alpha,
\label{4.3}
\end{equation}
then evaluates the Fourier transforms
\begin{equation}
\mbox{}_s \tilde{J}(\omega,\vec{x}) = \frac{1}{2\pi} \int
\mbox{}_s J(t,\vec{x})\, e^{i\omega t}\, dt,
\label{4.4}
\end{equation}
and takes the projections
\begin{equation}
\mbox{}_s \tilde{J}_{lm}(\omega;r) = \int 
\mbox{}_s \tilde{J}(\omega,\vec{x})\, 
\mbox{}_s \bar{Y}_{lm}(\theta,\phi)\, d\Omega.
\label{4.5}
\end{equation}
The source term is finally given by \cite{31}
\begin{equation}
T_{lm}(\omega; r) = 2\pi \sum_s \mbox{}_s p_l\,
\mbox{}_s {\cal D}\, \mbox{}_s \tilde{J}_{lm}(\omega;r),
\label{4.6}
\end{equation}
where the sum runs from $s=0$ to $s=-1$,
\begin{equation}
\mbox{}_s p_l = \left\{ 
\begin{array}{ll}
\sqrt{2l(l+1)} & \qquad s=0 \\
1 & \qquad s=-1
\end{array}
\right. ,
\label{4.7}
\end{equation}
and 
\begin{equation}
\mbox{}_s {\cal D} = \left\{
\begin{array}{ll}
r^3 & \qquad s=0 \\
r {\cal L} r^3 & \qquad s=-1
\end{array}
\right. ,
\label{4.8}
\end{equation}
with ${\cal L} = f d/dr + i\omega$. 

\subsection{Solution}

Equation (\ref{4.2}) is integrated by means of a Green's function,
in a manner similar to what was done in Sec.~III. We introduce two 
functions, $R^H_{l}(\omega;r)$ and $R^\infty_l(\omega;r)$, solutions 
to the homogeneous Teukolsky equation [Eq.~(\ref{4.2}) with 
$T_{lm}(\omega;r)=0$], with asymptotic behavior \cite{31}
\begin{eqnarray}
R^H_{l}(\omega; r\to 2M) &\sim& f e^{-i\omega r^*}, \nonumber \\
R^H_{l}(\omega; r\to\infty) &\sim& 
{\cal Q}^{\rm in}_l(\omega) (i\omega r)^{-1} e^{-i\omega r^*} + 
{\cal Q}^{\rm out}_l(\omega) (i\omega r) e^{i\omega r^*}, \label{4.9} \\
R^\infty_{l}(\omega; r\to\infty) &\sim& (i\omega r) e^{i\omega r^*}, 
\nonumber
\end{eqnarray}
where $r^* = r + 2M\ln(r/2M - 1)$. In terms of these, the
radial function is given by
\begin{eqnarray}
R_{lm}(\omega; r) &=& \frac{1}{2i\omega {\cal Q}^{\rm in}_l(\omega)}\,
\Biggl[ R^\infty_l(\omega;r) 
\int_{2M}^{r} \frac{T_{lm}(\omega;r') 
R^H_l(\omega;r')}{r'^2 f'} \, dr' 
\nonumber \\ & & \mbox{} 
+ R^H_l(\omega;r)
\int_{r}^{\infty} \frac{T_{lm}(\omega;r') 
R^\infty_l(\omega;r')}{r'^2 f'}\, dr' \Biggr],
\label{4.10}
\end{eqnarray}
where $f' = 1 -2M/r'$. 

As before, we are mostly interested in the behavior of the 
radial function near $r=\infty$. Equation (\ref{4.10}) gives
\begin{equation}
R_{lm}(\omega;r\to\infty) \sim \tilde{\cal Z}(\omega) (i\omega r)
e^{i\omega r^*}, 
\label{4.11}
\end{equation}
where
\begin{equation}
\tilde{\cal Z}_{lm}(\omega) \equiv 
\frac{1}{2i\omega {\cal Q}^{\rm in}_l(\omega)}\,
\int_{2M}^{\infty} \frac{T_{lm}(\omega;r) 
R^H_l(\omega;r)}{r^2 f} \, dr.
\label{4.12}
\end{equation}
These quantities are the (Fourier transform of the) radiative 
multipole moments of the electromagnetic field.

\subsection{Adjoint operators and Chandrasekhar transformation}

Equation (\ref{4.12}) will not be our final expression for the
radiative multipole moments. In Eq.~(\ref{4.12}), 
$T_{lm}(\omega;r)$ is obtained by applying the
differential operators $\mbox{}_s {\cal D}$ on 
$\mbox{}_s \tilde{J}_{lm}(\omega;r)$, the projections of
the current density. To have to take derivatives of these functions
is an inconvenience, and we would like to express the 
moments directly in terms of $\mbox{}_s \tilde{J}_{lm}(\omega;r)$. 
It is easy to show, by straightforward integration by parts, that if 
we define the adjoint operators
\begin{equation}
\mbox{}_s {\cal D}^\dagger = \left\{
\begin{array}{ll}
r^3 & \qquad s=0 \\
- r^5 \bar{\cal L} r^{-1} & \qquad s=-1
\end{array}
\right. ,
\label{4.13}
\end{equation}
where $\bar{\cal L} = f d/dr - i\omega$, then Eq.~(\ref{4.12})
is equivalent to
\begin{equation}
\tilde{\cal Z}_{lm}(\omega) = 
\frac{\pi}{i\omega {\cal Q}^{\rm in}_l(\omega)}\,
\sum_s \mbox{}_s p_l \int_{2M}^\infty 
\frac{\mbox{}_s \tilde{J}_{lm}(\omega;r) 
\mbox{}_s {\cal D}^\dagger R^H_l(\omega;r)}{r^2 f}\,
dr .
\label{4.14}
\end{equation}

Although $\tilde{\cal Z}_{lm}(\omega)$ is now expressed directly
in terms of $\mbox{}_s \tilde{J}_{lm}(\omega;r)$, Eq.~(\ref{4.14})
still will not be our final expression for the radiative
multipole moments. We now want to write $R^H_l(\omega;r)$ in
terms of $X^H_l(\omega;r)$, the solution to the 
generalized Regge-Wheeler equation (with $s=1$) considered 
in Sec.~II. The relationship between these functions was
trivial in the case of scalar radiation [cf.~Eq.~(\ref{3.17})].
That such a relationship exists in the case of gravitational 
radiation was shown by Chandrasekhar \cite{32}, who also provided it
explicitly. We show in Appendix E that for the case of electromagnetic
radiation, the Chandrasekhar transformation is given by 
\begin{equation}
\frac{R^H_l(\omega;r)}{{\cal Q}^{\rm in}_l(\omega)} = 
\frac{-2}{l(l+1)}\, r {\cal L}\,
\frac{X^H_l(\omega;r)}{{\cal A}^{\rm in}_l(\omega)}.
\label{4.15}
\end{equation}
Substituting this into Eq.~(\ref{4.14}), and taking into
account the fact that $\bar{\cal L} {\cal L} = l(l+1)f/r^2$
when acting on $X^H_l(\omega;r)$, we arrive at
\begin{equation}
\tilde{\cal Z}_{lm}(\omega) = 
\frac{-2\pi}{l(l+1)i\omega {\cal A}^{\rm in}_l(\omega)}\,
\sum_s \mbox{}_s p_l \int_{2M}^\infty 
\mbox{}_s \tilde{J}_{lm}(\omega;r)\,
\mbox{}_s \Gamma X^H_l(\omega;r)\, dr,
\label{4.16}
\end{equation}
where we have introduced the operators 
\begin{equation}
\mbox{}_s \Gamma = \left\{
\begin{array}{ll}
r^2 f^{-1} {\cal L} & \qquad s=0 \\
-l(l+1) r & \qquad s=-1
\end{array}
\right. .
\label{4.17}
\end{equation}
We emphasize that $X^H_l(\omega;r)$ denotes the solution to the 
$s=1$ generalized Regge-Wheeler equation with boundary conditions 
(\ref{2.4})--(\ref{2.5}). Equation (\ref{4.16}) {\it will} be our 
final expression for the radiative multipole moments. 

\subsection{Vector potential}

The physical meaning of the quantities $\tilde{Z}_{lm}(\omega)$
becomes more transparent if we use Eqs.~(\ref{4.1}) and (\ref{4.11}) 
to construct $A_\alpha^{\rm rad}(\bbox{x})$, the vector potential 
describing the radiative part of the field. [This is defined 
similarly to $\Phi_{\rm rad}(\bbox{x})$, encountered in Sec.~III.]

We may choose a gauge in which $A_\alpha^{\rm rad}(\bbox{x})$ 
is purely transverse to the direction of propagation, which, at
large distances from the source, is radially outward. This
implies that the vector potential may be expressed as 
\begin{equation}
A_\alpha^{\rm rad} = A\, m_\alpha + \bar{A}\, \bar{m}_\alpha,
\label{4.18}
\end{equation}
where $m_\alpha = (0,0,r,ir\sin\theta)/\sqrt{2}$ is complex
conjugate to $\bar{m}_\alpha$. It should be noted that the
quantity $A(\bbox{x})$ is complex, but that 
$A_\alpha^{\rm rad}(\bbox{x})$ is real.

The radiative part of the electromagnetic field tensor, 
$F_{\rm rad}^{\alpha\beta}(\bbox{x})$, can easily be computed 
from Eq.~(\ref{4.18}), and the Teukolsky field, $\Phi_2(\bbox{x}) = 
F_{\rm rad}^{\alpha\beta} \bar{m}_\alpha n_\beta$, follows
immediately. Keeping in mind that we are working near 
$r = \infty$, we find 
\begin{equation}
\Phi_2 = - A_{,\alpha} n^\alpha = - \frac{\partial A}{\partial u},
\label{4.19}
\end{equation}
where $u=t-r^*$. Combining this with Eqs.~(\ref{4.1}) and (\ref{4.11}) 
yields
\begin{equation}
A(t,\vec{x}) = \frac{1}{r}\, \sum_{lm} {\cal Z}_{lm}(u)\, 
\mbox{}_{-1} Y_{lm}(\theta,\phi),
\label{4.20}
\end{equation}
where
\begin{equation}
{\cal Z}_{lm}(u) = \int \tilde{Z}_{lm}(\omega)\, 
e^{i\omega u}\, d\omega.
\label{4.21}
\end{equation}
This shows that ${\cal Z}_{lm}(u)$ are indeed the radiative
multipole moments of the electromagnetic field. 

The vector potential is obtained by substituting Eq.(\ref{4.20})
into Eq.~(\ref{4.18}). The spin-weighted spherical harmonics then 
combine with the vectors $m_\alpha$ and $\bar{m}_\alpha$ to form 
the vectorial spherical harmonics described in Appendix A. [See
Eq.~(A16); Eq.~(\ref{A15}) must also be used.] We find
\begin{equation}
A^{\rm rad}_\alpha(t,\vec{x}) =  \frac{1}{r} \sum_{lm} 
\Bigl[ {\cal I}_{lm}(u)\, Y^{E,lm}_\alpha (\theta,\phi) +
{\cal S}_{lm}(u)\, Y^{B,lm}_\alpha (\theta,\phi) \Bigr],
\label{4.22}
\end{equation}
where we have introduced the {\it charge} multipole moments,
\begin{equation}
{\cal I}_{lm}(u) = \frac{1}{\sqrt{2}}\, \Bigl[ 
{\cal Z}_{lm}(u) + (-1)^m \bar{\cal Z}_{l,-m} (u) \Bigr],
\label{4.23}
\end{equation}
and the {\it current} multipole moments,
\begin{equation}
{\cal S}_{lm}(u) = \frac{i}{\sqrt{2}}\, \Bigl[ 
{\cal Z}_{lm}(u) - (-1)^m \bar{\cal Z}_{l,-m} (u) \Bigr].
\label{4.24}
\end{equation}
The reason for using this terminology will become clear below.
For the time being we may mention that the first group of terms 
in Eq.~(\ref{4.22}), that involving the charge moments, has 
electric-type parity, while the second group, involving the 
current moments, has magnetic-type parity (see Appendix A). 
The fact that two sets of multipole moments are needed to form 
$A^{\rm rad}_\alpha(\bbox{x})$ is related to the fact that the
electromagnetic field possesses two radiative degrees of freedom.

\subsection{Slow-motion approximation}

We now compute the radiative multipole moments (\ref{4.16})
in the slow motion approximation. The calculation is similar
to the one presented in Sec.~III C.

We begin by substituting Eq.~(\ref{2.20}), with $s=1$, into
Eq.~(\ref{4.16}). After using Eqs.~(\ref{4.3}), (\ref{4.5}), 
(\ref{4.7}), and (\ref{4.17}), we obtain the lengthy expression
\begin{eqnarray}
\tilde{\cal Z}_{lm}(\omega) &=& \frac{4\pi}{(2l+1)!!}\,
{\cal T}_l(\omega)\, (-i\omega)^l
\Biggl \lgroup \sqrt{\frac{l+1}{2l}} \int 
\biggl\{ 1 - \frac{l+3}{2(l+1)(2l+3)}\, (\omega r)^2 -
(l-1) \frac{M}{r} 
\nonumber \\ & & \mbox{}
+ \frac{i\omega r}{l+1} \biggl[ 1 - \frac{(\omega r)^2}{2(2l+3)} 
- \frac{l^2-2l-1}{l}\, \frac{M}{r} \biggr] + O(v^4) \biggr\}\, 
\Bigl( f \tilde{\rho}
+ \tilde{J}^r \Bigr)\,
r^l\, \mbox{}_0 \bar{Y}_{lm}\, d\vec{x} 
\nonumber \\ & & \mbox{}
- \int \biggl[ 1 - \frac{(\omega r)^2}{2(2l+3)} 
- \frac{l^2-1}{l}\, \frac{M}{r}
+ O(v^4) \biggr]\, \mbox{}_{-1} \tilde{J}\, 
r^l\, \mbox{}_{-1} \bar{Y}_{lm}\,
d\vec{x} \Biggr \rgroup.
\label{4.25}
\end{eqnarray}
Here, $\tilde{\rho}(\omega,\vec{x}) \equiv \tilde{J}^t(\omega,\vec{x})$ 
is the Fourier transform of the charge density, and
\begin{equation}
{\cal T}_l(\omega) = 
e^{2iM\omega (\ln 4M|\omega| - \beta_l)} 
\bigl( 1 + \pi M |\omega| \bigr),
\label{4.26}
\end{equation}
where $\beta_l$ is given by Eq.~(\ref{2.19}) with $s=1$:
\begin{equation}
\beta_l = \psi(l+1) + \frac{1}{2} - \frac{1}{2l(l+1)}.
\label{4.27}
\end{equation}

Inspecting Eq.~(\ref{4.25}), we notice that it does not have
the same mathematical structure as Eq.~(\ref{3.18}), which gives
the radiative multipole moments of a scalar field. In particular,
we see that $\tilde{\cal Z}_{lm}(\omega)$ possesses correction 
terms that are linear in $v$ [the terms 
$i\omega r \tilde{\rho} / (l+1)$ and $\tilde{J^r}$, the latter being
one power of $v$ smaller than $\tilde{\rho}$], as well as many 
third-order terms that depend explicitly on $r$, and which cannot
be taken outside the integral. This apparently contradicts our
expectation that $\tilde{\cal Z}_{lm}(\omega)$ should come with 
only near-zone corrections of order $v^2$, and wave-propagation
corrections of order $v^3$. 

However, expression (\ref{4.25}) is not unique, and we may use
the continuity equation $J^\alpha_{\ ;\alpha}= 0 $ to remove 
the unwanted terms. When written out explicitly, this reads
\begin{equation}
\rho_{,t} = - \frac{1}{r^2} \bigl(r^2 J^r \bigr)_{,r} - 
\frac{1}{\sqrt{2} r} \bigl( \hat{\partial} \mbox{}_{-1} J +
\check{\partial} \mbox{}_1 J \bigr),
\label{4.28}
\end{equation}
where $\hat{\partial}$ and $\check{\partial}$ are the ``edth''
differential operators described in Appendix A, and
\begin{equation}
\mbox{}_1 J = - J^\alpha m_\alpha.
\label{4.29}
\end{equation}
The continuity equation gives rise to an integral identity if
we multiply both sides by 
$r^n\, \mbox{}_0 \bar{Y}_{lm}(\theta,\phi)$ and integrate over
$d\vec{x}$. After a Fourier transform and several partial 
integrations [using Eqs.~(\ref{A13}) and (\ref{A14})], we obtain
\begin{equation}
-i\omega \int \tilde{\rho}\, r^n\, \mbox{}_0 \bar{Y}_{lm}\, d\vec{x} =
n \int \tilde{J}^r\, r^{n-1}\, \mbox{}_0 \bar{Y}_{lm}\, d\vec{x}
- \sqrt{\frac{l(l+1)}{2}} \int \Bigl( \mbox{}_{-1} \tilde{J}\,
\mbox{}_{-1} \bar{Y}_{lm} - \mbox{}_{1} \tilde{J}\,
\mbox{}_{1} \bar{Y}_{lm} \Bigr)\, r^{n-1}\, d\vec{x}.
\label{4.30}
\end{equation}
We now use this identity to remove all terms proportional to
$i\omega \tilde{\rho}(\omega,\vec{x})$ in Eq.~(\ref{4.25}). 
After some remarkable cancelations, wherein all unwanted terms
disappear, we arrive at our final expression for the radiative
multipole moments:
\begin{eqnarray}
\tilde{\cal Z}_{lm}(\omega) &=& \frac{4\pi}{(2l+1)!!}\,
{\cal T}_l(\omega)\, (-i\omega)^l
\Biggl\{ \sqrt{\frac{l+1}{2l}} \int 
\biggl[ 1 
- \frac{l+3}{2(l+1)(2l+3)}\, (\omega r)^2 - (l-1) \frac{M}{r}
\nonumber \\ & & \mbox{}
+ O(v^4) \biggr] 
\biggl( \tilde{\rho} + \frac{i\omega r}{l+1}\, \tilde{J}^r \biggr)\,
r^l\, \mbox{}_{0} \bar{Y}_{lm}\, d\vec{x}
- \frac{1}{2} \int \biggl[ 1 - \frac{(\omega r)^2}{2(2l+3)} 
- \frac{l^2-1}{l}\, \frac{M}{r}
\nonumber \\ & & \mbox{}
+ O(v^4) \biggr]\,
\Bigl( \mbox{}_{-1} \tilde{J} \mbox{}_{-1} \bar{Y}_{lm} + 
\mbox{}_{1} \tilde{J} \mbox{}_{1} \bar{Y}_{lm} \Bigr)\, 
 r^l\, d\vec{x} \Biggr\}.
\label{4.31}
\end{eqnarray}
We see that this expression has the expected form, with all
$O(v^2)$ corrections occurring inside the spatial integrals, and 
all $O(v^3)$ corrections occurring outside.

We now separate $\tilde{\cal Z}_{lm}(\omega)$ into charge and current 
moments, according to the Fourier transform of Eqs.~(\ref{4.23}) and 
(\ref{4.24}). This gives
\begin{eqnarray}
\tilde{\cal I}_{lm}(\omega) &=&  \frac{4\pi}{(2l+1)!!}\,
\sqrt{\frac{l+1}{l}}\, {\cal T}_l(\omega)\, (-i\omega)^l
\int \biggl[ 1 
- \frac{l+3}{2(l+1)(2l+3)}\, (\omega r)^2 
\nonumber \\ & & \mbox{}
- (l-1) \frac{M}{r} + O(v^4) \biggr]
\biggl[ \tilde{\rho}(\omega,\vec{x})
+ \frac{i\omega r}{l+1}\, \tilde{J}^r(\omega,\vec{x}) \biggr]\,
r^l\, \mbox{}_{0} \bar{Y}_{lm}(\theta,\phi)\, d\vec{x}
\label{4.32}
\end{eqnarray}
and
\begin{eqnarray}
\tilde{\cal S}_{lm}(\omega) &=& - \frac{2\sqrt{2}i\pi}{(2l+1)!!}\,
{\cal T}_l(\omega)\, (-i\omega)^l
\int \biggl[ 1 - \frac{(\omega r)^2}{2(2l+3)} 
- \frac{l^2-1}{l}\, \frac{M}{r}
\nonumber \\ & & \mbox{}
+ O(v^4) \biggr]
\biggl[ \mbox{}_{-1} \tilde{J}(\omega,\vec{x}) 
\mbox{}_{-1} \bar{Y}_{lm}(\theta,\phi) + 
\mbox{}_{1} \tilde{J}(\omega,\vec{x}) 
\mbox{}_{1} \bar{Y}_{lm}(\theta,\phi) \biggr]\, 
 r^l\, d\vec{x}.
\label{4.33}
\end{eqnarray}
The corresponding expressions in the time domain (see Appendix D) are
\begin{equation}
{\cal I}_{lm}(u) = I_{lm}(u) + 2M \int_{-\infty}^u \biggl[
\ln \biggl( \frac{u-u'}{4M} \biggr) + \beta_l + \gamma \biggr]\,
\ddot{I}_{lm}(u')\, du'
\label{4.34}
\end{equation}
and
\begin{equation}
{\cal S}_{lm}(u) = S_{lm}(u) + 2M \int_{-\infty}^u \biggl[
\ln \biggl( \frac{u-u'}{4M} \biggr) + \beta_l + \gamma \biggr]\,
\ddot{S}_{lm}(u')\, du',
\label{4.35}
\end{equation}
where a dot indicates differentiation with respect to $u'$. We
have defined
\begin{eqnarray}
I_{lm}(u) &=& \frac{4\pi}{(2l+1)!!}\, \sqrt{\frac{l+1}{l}} 
\biggl( \frac{d}{du} \biggr)^l \int 
\biggl[ 1 + \frac{l+3}{2(l+1)(2l+3)}\, (r \partial_u)^2 
\nonumber \\ & & \mbox{} - 
(l-1) \frac{M}{r} + O(v^4) \biggr]
\biggl[ \rho(u,\vec{x}) - 
\frac{r \partial_u}{l+1}\, J^r(u,\vec{x}) \biggr]
\, r^l\, \mbox{}_{0} \bar{Y}_{lm}(\theta,\phi)\, d\vec{x}
\label{4.36}
\end{eqnarray}
and
\begin{eqnarray}
S_{lm}(u) &=& -\frac{2\sqrt{2}i\pi}{(2l+1)!!}\,
\biggl( \frac{d}{du} \biggr)^l 
\int \Biggl[ 1 + \frac{(r \partial_u)^2}{2(2l+3)} - 
\frac{l^2-1}{l}\, \frac{M}{r} + O(v^4) \biggr]
\nonumber \\ & & \mbox{} \times 
\biggl[ \mbox{}_{-1} J(u,\vec{x})\, 
\mbox{}_{-1} \bar{Y}_{lm}(\theta,\phi) + 
\mbox{}_{1} J(u,\vec{x})\, 
\mbox{}_{1} \bar{Y}_{lm}(\theta,\phi) \biggr]\,
r^l\, d\vec{x}.
\label{4.37}
\end{eqnarray}
Equations (\ref{4.34})--(\ref{4.37}) have the same mathematical 
structure as Eqs.~(\ref{3.20}) and (\ref{3.21}), which give the 
radiative multipole moments of a scalar field. The physical meaning
of these equations is therefore exactly the same as in Sec.~III, and
the discussion appearing at the end of Sec.~III E need not be 
repeated. Equations (\ref{4.34})--(\ref{4.37}), with $\beta_l$
given by Eq.~(\ref{4.27}), are equivalent to 
Eqs.~(\ref{1.8})--(\ref{1.12}), once the spin-weighted spherical
harmonics have been converted into the vectorial harmonics of 
Eq.~(\ref{A16}).

\section{Gravitational radiation}

\subsection{Teukolsky equation}

In this section, we derive expressions for the radiative multipole 
moments of a gravitational-wave field in Schwarzschild spacetime. 
Specifically, we consider a tensor field $h_{\alpha\beta}(\bbox{x})$, 
defined as the difference between the metric of the perturbed spacetime
and the Schwarzschild metric. Field equations for 
$h_{\alpha\beta}(\bbox{x})$ are obtained by linearizing
the Einstein equations for the full metric. It is therefore 
assumed that the perturbation is small. Teukolsky \cite{31} 
has cast the field equations for $h_{\alpha\beta}(\bbox{x})$ in a 
form convenient for our purposes. We briefly summarize this formulation 
here. 

In the Teukolsky formalism, the radiative part of 
$h_{\alpha\beta}(\bbox{x})$ is represented by the 
complex-valued function $\Psi_4 = -C_{\alpha\beta\gamma\delta} 
n^\alpha \bar{m}^\beta n^\gamma \bar{m}^\delta$, where
$C_{\alpha\beta\gamma\delta}$ is the perturbed Weyl tensor, and
$n^\alpha$ and $\bar{m}^\alpha$ are defined as in Sec.~IV. 
The field $\Psi_4(\bbox{x})$ has spin-weight $s=-2$ (see Appendix A), 
and it can be decomposed according to
\begin{equation}
\Psi_4 = \frac{1}{r^4} \int d \omega \sum_{lm} R_{lm} (\omega;r)\,
\mbox{}_{-2} Y_{l m} (\theta,\phi)\, e^{-i \omega t},
\label{5.1}
\end{equation}
where $\mbox{}_{-2} Y_{l m} (\theta,\phi)$ are spin-weighted
spherical harmonics (see Appendix A). The sums over $l$ and 
$m$ are restricted by $l \geq 2$ and 
$|m| \leq l$. The radial function then satisfies the
inhomogeneous Teukolsky equation \cite{31},
\begin{equation}
\Biggl[ r^2 f \frac{d^2}{dr^2} - 2(r-M) \frac{d}{dr}
+ U(\omega;r) \Biggr] R_{lm}(\omega;r) = T_{lm}(\omega;r),
\label{5.2}
\end{equation}
where $f=1-2M/r$ and $U(\omega;r) = f^{-1}[ (\omega r)^2 - 
4i\omega (r-3M)] - (l-1)(l+2)$.

The source term to the right-hand side of Eq.~(\ref{5.2}) is 
constructed as follows from $T^{\alpha\beta}(\bbox{x})$, the 
(unspecified) stress-energy tensor responsible for the 
perturbation. The first step is to form the contractions
\begin{equation}
\mbox{}_0 T = T_{\alpha\beta} n^\alpha n^\beta, \qquad
\mbox{}_{-1} T = T_{\alpha\beta} n^\alpha \bar{m}^\beta, \qquad
\mbox{}_{-2} T = T_{\alpha\beta} \bar{m}^\alpha \bar{m}^\beta.
\label{5.3}
\end{equation}
One then evaluates the Fourier transforms
\begin{equation}
\mbox{}_s \tilde{T}(\omega,\vec{x}) = \frac{1}{2\pi}
\int \mbox{}_s T(t,\vec{x})\, e^{i\omega t}\, dt
\label{5.4}
\end{equation}
and takes the projections
\begin{equation}
\mbox{}_s \tilde{T}_{lm} (\omega;r) = \int 
\mbox{}_s \tilde{T}(\omega,\vec{x})\, \mbox{}_s \bar{Y}_{l m}
(\theta,\phi)\, d\Omega.
\label{5.5}
\end{equation}
Finally, $T_{lm}(\omega;r)$ is given by \cite{31}
\begin{equation}
T_{lm}(\omega;r) = 2\pi \sum_s \mbox{}_s p_l\, 
\mbox{}_s {\cal D}\, \mbox{}_s \tilde{T}_{lm}(\omega;r),
\label{5.6}
\end{equation}
where 
\begin{equation}
\mbox{}_s p_l = \left\{
\begin{array}{ll}
2 \sqrt{(l-1)l(l+1)(l+2)} & \qquad s=0 \\
2 \sqrt{2(l-1)(l+2)} & \qquad s=-1 \\
1 & \qquad s=-2  
\end{array} \right. 
\label{5.7}
\end{equation}
and
\begin{equation}
\mbox{}_s {\cal D} = \left\{
\begin{array}{ll}
r^4 & \qquad s=0 \\
r^2 f {\cal L} r^3 f^{-1} & \qquad s=-1 \\
r f {\cal L} r^4 f^{-1} {\cal L} r & \qquad s=-2
\end{array}
\right. . 
\label{5.8}
\end{equation}
Here, ${\cal L} = f d/dr + i\omega$.

\subsection{Solution}

The inhomogeneous Teukolsky equation (\ref{5.2}) can be
integrated by means of a Green's function, in a manner
similar to what was done in Sec.~IV. Here also, the form
of the radial function can be simplified by introducing adjoint 
operators $\mbox{}_s {\cal D}^\dagger$, and by expressing 
it in terms of $X^H_l(\omega;r)/{\cal A}^{\rm in}_l(\omega)$, 
where $X^H_l(\omega;r)$ is the solution to the Regge-Wheeler 
equation --- Eq.~(\ref{2.1}) with $s=2$ --- with boundary
conditions (\ref{2.4}), (\ref{2.5}). These manipulations
are described in detail in Ref.~\cite{28}, and they
will not be displayed here. The conclusion is that at large
distances, the radial function is given by
\begin{equation}
R_{lm}(\omega;r\to\infty) \sim \frac{1}{2}\, \omega^2\,
\tilde{\cal Z}_{lm}(\omega)\, r^3 e^{i\omega r^*},
\label{5.9}
\end{equation}
where
\begin{equation}
\tilde{\cal Z}_{lm}(\omega) = 
\frac{-2 \pi}{i \omega \kappa_l(\omega) {\cal A}^{\rm in}_l(\omega)}\,  
\sum_s \mbox{}_s p_l \int_{2M}^\infty 
r f^{-2}\, \mbox{}_s \tilde{T}_{lm}(\omega; r)\, 
\mbox{}_s \Gamma\, X^H_{l} (\omega;r).
\label{5.10}
\end{equation}
Here, 
\begin{equation}
\kappa_l(\omega) = \frac{1}{4} \Bigl[
(l-1)l(l+1)(l+2) - 12 i M \omega \Bigr],
\label{5.11}
\end{equation}
and 
\begin{eqnarray}
\mbox{}_0 \Gamma &=& 
2(1-3M/r+i\omega r) r f \frac{d}{dr} 
+ f \bigl[ l(l+1) - 6M/r \bigr]
+ 2i\omega r (1 - 3M/r + i\omega r), 
\nonumber \\
\mbox{}_{-1} \Gamma &=& -f \Bigl\{
\bigl[ l(l+1)+2i\omega r \bigr] r f \frac{d}{dr}
+ l(l+1)(f+i\omega r) - 2 (\omega r)^2 \Bigr\}, 
\label{5.12} \\
\mbox{}_{-2} \Gamma &=& f^2 \Bigl\{
2 \bigl[ (l-1)(l+2)+6M/r \bigr] rf \frac{d}{dr}
+ (l-1)(l+2)\bigl[l(l+1) + 2i\omega r \bigr] 
+ 12 f M/r \Bigr\}. 
\nonumber
\end{eqnarray}
The quantities $\tilde{\cal Z}_{lm}(\omega)$ 
are the multipole moments of the radiative part of 
$h_{\alpha\beta}(\bbox{x})$.

\subsection{Metric perturbation}

The gravitational-wave field, $h^{\rm rad}_{\alpha\beta}(\bbox{x})$, 
can be obtained from the behavior of $\Psi_4(\bbox{x})$ at 
large distances \cite{28}. Choosing the $\theta$ and $\phi$ 
directions as polarization axes, the two fundamental polarizations 
of the gravitational waves are given by 
\begin{equation}
h_+(t,\vec{x}) - i h_\times(t,\vec{x}) = 
\frac{1}{r} \sum_{l m} {\cal Z}_{lm}(u)\,
\mbox{}_{-2} Y_{lm}(\theta,\phi), 
\label{5.13}
\end{equation}
where $u=t-r^*$, and
\begin{equation}
{\cal Z}_{lm} (u) = \int_{-\infty}^\infty 
\tilde{\cal Z}_{lm}(\omega)\, e^{-i \omega u}\, d\omega.
\label{5.14}
\end{equation}
This shows that ${\cal Z}_{lm}(u)$ are indeed the multipole
moments of the radiative field. It should be noted that while
these quantities are complex, $h_+(\bbox{x})$ and 
$h_\times(\bbox{x})$ are real.

In the traceless-transverse gauge, the gravitational-wave tensor 
is given by
\begin{equation}
h^{\rm rad}_{\alpha\beta} = (h_+ - ih_\times)\, m_\alpha m_\beta + 
(h_+ + i h_\times)\, \bar{m}_\alpha \bar{m}_\beta,
\label{5.15}
\end{equation}
or, after substituting Eq.~(\ref{5.13}),
\begin{equation}
h^{\rm rad}_{\alpha\beta}(t,\vec{x}) = \frac{1}{r} \sum_{lm} 
\Bigl[ {\cal I}_{lm}(u)\, T^{E2,lm}_{\alpha\beta}(\theta,\phi) 
+ {\cal S}_{lm}(u)\, T^{B2,lm}_{\alpha\beta}(\theta,\phi) \Bigr].
\label{5.16}
\end{equation}
Here, $h^{\rm rad}_{\alpha\beta}(\bbox{x})$ is expressed in terms
of the tensorial spherical harmonics described in Appendix A. 
The {\it mass} multipole moments ${\cal I}_{lm}(u)$,
and the {\it current} multipole moments ${\cal S}_{lm}(u)$, are
related to ${\cal Z}_{lm}(u)$ by the same equations as 
Eqs.~(\ref{4.23}) and (\ref{4.24}). 

\subsection{Slow-motion approximation}

We now calculate the radiative multipole moments in the 
slow-motion approximation. We proceed as in Sec.~IV E. 
Substituting Eq.~(\ref{2.20}), with $s=2$, into 
Eq.~(\ref{5.10}), and using Eqs.~(\ref{5.5}), (\ref{5.7}),
(\ref{5.11}) and (\ref{5.12}), yields
\begin{equation}
\tilde{\cal Z}_{lm}(\omega) = \frac{16\pi}{(2l+1)!!}\,
{\cal T}_l(\omega)\, (-i\omega)^l\, 
\sum_s \mbox{}_s \tilde{P}_{lm}(\omega),
\label{5.17}
\end{equation}
where
\begin{equation}
{\cal T}_l(\omega) = 
e^{2iM\omega (\ln 4M|\omega| - \mu_l)} 
\bigl( 1 + \pi M |\omega| \bigr),
\label{5.18}
\end{equation}
with
\begin{equation}
\mu_l = \beta_l - \frac{6}{(l-1)l(l+1)(l+2)},
\label{5.19}
\end{equation}
while $\beta_l$ is given by Eq.~(\ref{2.19}) with $s=2$:
\begin{equation}
\beta_l = \psi(l+1) + \frac{1}{2} - \frac{2}{l(l+1)}.
\label{5.20}
\end{equation}
We have also introduced
\begin{eqnarray}
\mbox{}_0 \tilde{P}_{lm}(\omega) &=& 2 \sqrt{\frac{(l+1)(l+2)}{(l-1)l}}\,
\int \biggl\{ 1 - \frac{l+9}{2(l+1)(2l+3)}\, (\omega r)^2 - 
(l+2) \frac{M}{r} 
+ \frac{2i\omega r}{l+1} \biggl[ 1 
\nonumber \\ & & \mbox{} 
- \frac{l+4}{2(l+2)(2l+3)}\, (\omega r)^2 - 
\frac{l^3+3l^2+l-4}{l(l+2)}\, \frac{M}{r} \biggr] 
+ O(v^4) \biggr\}\, \frac{\mbox{}_0 \tilde{T}}{f^2}\,
r^l\, \mbox{}_{0} \bar{Y}_{lm}\, d\vec{x}, 
\label{5.21} \\
\mbox{}_{-1} \tilde{P}_{lm}(\omega) &=& - 2 \sqrt{\frac{2(l+2)}{l-1}}\,
\int \biggl\{ 1 - \frac{l^2+3l+6}{2l(l+1)(2l+3)}\, (\omega r)^2 - 
\frac{(l-1)(l+2)}{l}\, \frac{M}{r} 
+ \frac{i\omega r}{l} \biggl[ 1 
\nonumber \\ & & \mbox{} 
- \frac{l^2+3l+6}{2(l+1)(l+2)(2l+3)}\, (\omega r)^2 - 
\frac{l^3+3l^2-8}{(l+1)(l+2)}\, \frac{M}{r} \biggr] 
+ O(v^4) \biggr\}\, \frac{\mbox{}_{-1} \tilde{T}}{f}\,
r^l\, \mbox{}_{-1} \bar{Y}_{lm}\, d\vec{x}, 
\label{5.22} \\
\mbox{}_{-2} \tilde{P}_{lm}(\omega) &=& \frac{l+2}{l}\,
\int \biggl[ 1 + \frac{2 i\omega r}{(l+1)(l+2)} + O(v^2)
\biggr]\, \mbox{}_{-2} \tilde{T}\,
r^l\, \mbox{}_{-2} \bar{Y}_{lm}\, d\vec{x}. 
\label{5.23}
\end{eqnarray}

To better keep track of the relative importance of each term in 
Eq.~(\ref{5.17}), we decompose the stress-energy tensor according to
\begin{eqnarray}
& \rho = T^{tt}, & 
\label{5.24} \\
& \mbox{}_{0} j = T^{tr}, \qquad 
\mbox{}_{-1} j = - T^{t\alpha} \bar{m}_{\alpha}, \qquad
\mbox{}_{1} j = - T^{t\alpha} m_{\alpha}, & 
\label{5.25} \\
& \mbox{}_{0} p = T^{rr}, \qquad 
\mbox{}_{-1} p = - T^{r\alpha} \bar{m}_{\alpha}, \qquad
\mbox{}_{1} p = - T^{r\alpha} m_{\alpha}, &
\label{5.26} \\
& \mbox{}_{0} t = T^{\alpha\beta} m_\alpha \bar{m}_\beta, \qquad
\mbox{}_{-2} t = T^{\alpha\beta} \bar{m}_\alpha \bar{m}_\beta, \qquad
\mbox{}_{2} t = T^{\alpha\beta} m_\alpha m_\beta. &
\label{5.27}
\end{eqnarray}
Thus, if $\rho$ is considered to be a quantity of order unity, then
$\mbox{}_s j = O(v)$, $\mbox{}_s p = O(v^2)$, and $\mbox{}_s t =
O(v^2)$. Inspection of Eqs.~(\ref{5.21})--(\ref{5.23}) --- in
which we substitute $\mbox{}_0 T/f^2 = \frac{1}{4} (
\rho + 2 \mbox{}_0 j/f + \mbox{}_0 p/f^2)$,
$\mbox{}_{-1} T/f = \frac{1}{2} (\mbox{}_{-1} j +
\mbox{}_{-1} p / f)$, and $\mbox{}_{-2} T = 
\mbox{}_{-2} t$ --- then reveals that as it stands, 
$\tilde{\cal Z}_{lm}(\omega)$ contains many unwanted terms 
of the sort encountered in Sec.~IV E: terms which are first order 
in $v$, and $O(v^3)$ terms that cannot be pulled outside of the 
spatial integrals. 

In Sec.~IV E, the unwanted terms were removed by invoking the
continuity equation, $J^{\alpha}_{\ ;\alpha}=0$. Here, they are
removed with the help of the energy-momentum conservation 
equations, $T^{\alpha\beta}_{\ \ \ ;\beta} = 0$. When written
out explicitly, these become
\begin{eqnarray}
& \displaystyle
\rho_{,t} + \frac{1}{r^2 f} \bigl( r^2 f \mbox{}_0 j \bigr)_{,r}
+ \frac{1}{\sqrt{2}r} \bigl( \hat{\partial} \mbox{}_{-1} j + 
\check{\partial} \mbox{}_{1} j \bigr) = 0, &
\label{5.28} \\
& \displaystyle
\mbox{}_0 j_{,t} + \frac{\sqrt{f}}{r^2} \biggl( \frac{r^2}{\sqrt{f}}\,
\mbox{}_0 p \biggr)_{,r} + \frac{1}{\sqrt{2}r} 
\bigl( \hat{\partial} \mbox{}_{-1} p + 
\check{\partial} \mbox{}_{1} p \bigr) + \frac{1}{2} f f' \rho
- \frac{2f}{r}\, \mbox{}_0 t = 0, &
\label{5.29} \\
& \displaystyle
\mbox{}_{-1} j_{,t} + \frac{1}{r^3} 
\bigl( r^3 \mbox{}_{-1} p \bigr)_{,r} + 
\frac{1}{\sqrt{2}r} \bigl( \hat{\partial} \mbox{}_{-2} t + 
\check{\partial} \mbox{}_{0} t \bigr) = 0, &
\label{5.30} \\
& \displaystyle
\mbox{}_{1} j_{,t} + \frac{1}{r^3} 
\bigl( r^3 \mbox{}_{1} p \bigr)_{,r} + 
\frac{1}{\sqrt{2}r} \bigl( \hat{\partial} \mbox{}_{0} t + 
\check{\partial} \mbox{}_{-2} t \bigr) = 0, &
\label{5.31}
\end{eqnarray}
where $\hat{\partial}$ and $\check{\partial}$ are the ``edth''
differential operators described in Appendix A, and $f' = df/dr
= 2M/r^2$.

Equations (\ref{5.28})--(\ref{5.31}) give rise to a number of integral 
identities, which we write in the frequency domain, and which are easily
established by partial integration, using Eqs.~(\ref{A13}) and (\ref{A14}). 
We shall need the following two identities, which involve no approximation:
\begin{eqnarray}
i\omega \int \tilde{\rho}\, r^n\, 
\mbox{}_{0} \bar{Y}_{lm}\, d\vec{x} &=& 
- \int (n -r f'/f) \mbox{}_{0} \tilde{\jmath}\,
r^{n-1}\, \mbox{}_{0} \bar{Y}_{lm}\, d\vec{x}
\nonumber \\ & & \mbox{} 
+ \sqrt{ \frac{l(l+1)}{2} } \int \bigl( \mbox{}_{-1} \tilde{\jmath}\,
\mbox{}_{-1} \bar{Y}_{lm} - \mbox{}_{1} \tilde{\jmath}\,
\mbox{}_{1} \bar{Y}_{lm} \bigr)\, r^{n-1}\, d\vec{x}
\label{5.32}
\end{eqnarray}
and
\begin{eqnarray}
i\omega \int \mbox{}_{-1} \tilde{\jmath}\, r^n\, 
\mbox{}_{-1} \bar{Y}_{lm}\, d\vec{x} &=& 
- (n-1) \int \mbox{}_{-1} \tilde{p}\, r^{n-1}\, 
\mbox{}_{-1} \bar{Y}_{lm}\, d\vec{x} 
+ \sqrt{ \frac{(l-1)(l+2)}{2} } \int 
\mbox{}_{-2} \tilde{t}\, r^{n-1}\, 
\mbox{}_{-2} \bar{Y}_{lm}\, d\vec{x}
\nonumber \\ & & \mbox{} 
- \sqrt{ \frac{l(l+1)}{2} } \int 
\mbox{}_{0} \tilde{t}\, r^{n-1}\, 
\mbox{}_{0} \bar{Y}_{lm}\, d\vec{x}.
\label{5.33}
\end{eqnarray}
We shall also need the following identities, which are valid in the
slow-motion approximation:
\begin{eqnarray}
\int \bigl[ (\omega r)^2 - lM/r \bigr]\, 
\mbox{}_0 \tilde{\jmath}\, r^l\, \mbox{}_{0} \bar{Y}_{lm}\, d\vec{x} &=&
- \sqrt{\frac{l(l+1)}{2}} \int  \frac{M}{r}\,
\bigl( \mbox{}_{-1}\tilde{\jmath}\, \mbox{}_{-1} \bar{Y}_{lm} 
- \mbox{}_{1}\tilde{\jmath}\, \mbox{}_{1} \bar{Y}_{lm} \bigr)\,
r^l\, d\vec{x} 
\nonumber \\ & & \mbox{} 
+ \int i\omega r\,  
\bigl[ (l+2) \mbox{}_0 \tilde{p} + 2 \mbox{}_0 \tilde{t} \bigr]\,
r^l\, \mbox{}_0 \bar{Y}_{lm}\, d\vec{x} 
\nonumber \\ & & \mbox{} 
- \sqrt{\frac{l(l+1)}{2}} \int i\omega r\,
\bigl( \mbox{}_{-1}\tilde{p}\, \mbox{}_{-1} \bar{Y}_{lm} -
\mbox{}_{1}\tilde{p}\, \mbox{}_{1} \bar{Y}_{lm} \bigr)\,
r^l\, d\vec{x},
\label{5.34}
\end{eqnarray}
where terms of order $v^5$ and higher (with $\tilde{\rho}$
taken to be of order unity) have been discarded, and
\begin{equation}
\int \bigl[ iM\omega + O(v^5) \bigr]\, \mbox{}_{-1} \tilde{\jmath}\,
r^l\, \mbox{}_{-1} \bar{Y}_{lm}\, d\vec{x} = -(l-1) \int
\bigl[ M/r + O(v^4) \bigr]\, \mbox{}_{-1} \tilde{p}\,
r^l\, \mbox{}_{-1} \bar{Y}_{lm}\, d\vec{x},
\label{5.35}
\end{equation}
which follows from Eq.~(\ref{5.33}) after multiplying both
sides by $M$. 

These identities are used to remove all unwanted 
terms from $\tilde{\cal Z}_{lm}(\omega)$, as given by 
Eqs.~(\ref{5.17})--(\ref{5.23}). After a rather long calculation
(which spans several pages), we arrive at the 
following expression for the radiative multipole moments:
\begin{eqnarray}
\tilde{\cal Z}_{lm}(\omega) &=& \frac{8\pi}{(2l+1)!!}\, 
{\cal T}_l(\omega)\, (-i\omega)^l \Biggl\{
\sqrt{ \frac{(l+1)(l+2)}{(l-1)l} } \int \biggl[
1 - \frac{l+9}{2(l+1)(2l+3)}\, (\omega r)^2 
\nonumber \\ & & \mbox{} 
- (l+2)\frac{M}{r}
+ O(v^4) \biggr] \biggl( \tilde{\rho} + \mbox{}_0 \tilde{p}
+ 2 \mbox{}_0 \tilde{t} + \frac{4i\omega r}{l+1}\, \mbox{}_0 
\tilde{\jmath} \biggr)\, r^l\, \mbox{}_0 \bar{Y}_{lm}\, d\vec{x} 
- \sqrt{ \frac{2(l+2)}{l-1} } \int \biggl[ 1 
\nonumber \\ & & \mbox{} 
- \frac{l+4}{2(l+2)(2l+3)}\, (\omega r)^2 - 
\frac{(l-1)(l+2)}{l}\, \frac{M}{r} - 
\frac{24 i M\omega }{(l-1)l(l+1)(l+2)} 
\nonumber \\ & & \mbox{} 
+ O(v^4) \biggr]
\biggl( \mbox{}_{-1} \tilde{\jmath} + \frac{i\omega r}{l+2}\,
\mbox{}_{-1} \tilde{p} \biggr)\, r^l\, \mbox{}_{-1} \bar{Y}_{lm}\, 
d\vec{x} - \sqrt{ \frac{2(l+2)}{l-1} } \int \biggl[
1 - \frac{l+4}{2(l+2)(2l+3)}\, (\omega r)^2  
\nonumber \\ & & \mbox{} 
- \frac{(l-1)(l+2)}{l}\, \frac{M}{r} + O(v^4) \biggr]
\biggl( \mbox{}_{1} \tilde{\jmath} + \frac{i\omega r}{l+2}\,
\mbox{}_{1} \tilde{p} \biggr)\, r^l\, \mbox{}_{1} \bar{Y}_{lm}\, 
d\vec{x} \Biggr\}.
\label{5.36}
\end{eqnarray}
Notice that the second and third integrals differ by a term proportional 
to $iM\omega$. We see that this expression for the radiative 
multipole moments has the expected form, with all $O(v^2)$
correction terms occurring inside the spatial integrals, and all
$O(v^3)$ corrections occurring outside.

We now separate $\tilde{\cal Z}_{lm}(\omega)$ into mass and current
moments, according to the Fourier transform of Eqs.~(\ref{4.23}) and 
(\ref{4.24}). (As was pointed out in Sec.~IV C, those equations are
valid also in the case of gravitational radiation.) We find that
the mass moments are given by 
\begin{eqnarray}
\tilde{\cal I}_{lm}(\omega) &=& \frac{16\pi}{(2l+1)!!}\, 
\sqrt{ \frac{(l+1)(l+2)}{2(l-1)l} }\,
{\cal T}_l(\omega)\, (-i\omega)^l \int \biggl[
1 - \frac{l+9}{2(l+1)(2l+3)}\, (\omega r)^2 
- (l+2)\frac{M}{r}
\nonumber \\ & & \mbox{} 
+ O(v^4) \biggr] \biggl[ \tilde{\rho}(\omega,\vec{x}) + 
\mbox{}_0 \tilde{p}(\omega,\vec{x}) + 
2 \mbox{}_0 \tilde{t}(\omega,\vec{x}) + 
\frac{4i\omega r}{l+1}\, \mbox{}_0 \tilde{\jmath}(\omega,\vec{x}) 
\biggr]\, r^l\, \mbox{}_0 \bar{Y}_{lm}(\theta,\phi)\, d\vec{x},
\label{5.37}
\end{eqnarray}
and that the current moments are given by
\begin{eqnarray}
\tilde{\cal S}_{lm}(\omega) &=& - \frac{16\pi i}{(2l+1)!!}\,
\sqrt{ \frac{l+2}{l-1} }\, {\cal T}^\sharp_l(\omega)\, 
(-i\omega)^l \int \biggl[ 
1 - \frac{l+4}{2(l+2)(2l+3)}\, (\omega r)^2 
\nonumber \\ & & \mbox{} 
- \frac{(l-1)(l+2)}{l}\, \frac{M}{r} + O(v^4) \biggr]
\biggl\{ \biggl[ \mbox{}_{-1} \tilde{\jmath}(\omega,\vec{x}) + 
\frac{i\omega r}{l+2}\, \mbox{}_{-1} \tilde{p}(\omega,\vec{x}) 
\biggr]\, \mbox{}_{-1} \bar{Y}_{lm}(\theta,\phi)  
\nonumber \\ & & \mbox{} 
+ \biggl[ \mbox{}_{1} \tilde{\jmath}(\omega,\vec{x}) + 
\frac{i\omega r}{l+2}\, \mbox{}_{1} \tilde{p}(\omega,\vec{x}) 
\biggr]\, \mbox{}_{1} \bar{Y}_{lm}(\theta,\phi) \biggr\}\,
r^l\, d\vec{x},
\label{5.38}
\end{eqnarray}
where
\begin{equation}
{\cal T}^\sharp_l(\omega) = 
e^{2iM\omega (\ln 4M|\omega| - \beta_l)} 
\bigl( 1 + \pi M |\omega| \bigr).
\label{5.39}
\end{equation}
Notice that different constants ($\mu_l$ for the mass moments, 
$\beta_l$ for the current moments) appear in ${\cal T}_l(\omega)$ and 
${\cal T}^\sharp_l(\omega)$; these are defined by Eqs.~(\ref{5.19})
and (\ref{5.20}). 

The corresponding expressions in the time domain (see Appendix D) are
\begin{equation}
{\cal I}_{lm}(u) = I_{lm}(u) + 2M \int_{-\infty}^u \biggl[
\ln \biggl( \frac{u-u'}{4M} \biggr) + \mu_l + \gamma \biggr]\,
\ddot{I}_{lm}(u')\, du'
\label{5.40}
\end{equation}
and
\begin{equation}
{\cal S}_{lm}(u) = S_{lm}(u) + 2M \int_{-\infty}^u \biggl[
\ln \biggl( \frac{u-u'}{4M} \biggr) + \beta_l + \gamma \biggr]\,
\ddot{S}_{lm}(u')\, du',
\label{5.41}
\end{equation}
where a dot indicates differentiation with respect to $u'$. We
have defined
\begin{eqnarray}
I_{lm}(u) &=& \frac{16\pi}{(2l+1)!!}\, 
\sqrt{ \frac{(l+1)(l+2)}{2(l-1)l} }\,
\biggl( \frac{d}{du} \biggr)^l
\int \biggl[
1 + \frac{l+9}{2(l+1)(2l+3)}\, (r \partial_u)^2 
- (l+2)\frac{M}{r}
\nonumber \\ & & \mbox{} 
+ O(v^4) \biggr] \biggl[ \rho(u,\vec{x}) + 
\mbox{}_0 p(u,\vec{x}) + 
2 \mbox{}_0 t(u,\vec{x}) - 
\frac{4 r \partial_u}{l+1}\, \mbox{}_0 j(u,\vec{x}) 
\biggr]\, r^l\, \mbox{}_0 \bar{Y}_{lm}(\theta,\phi)\, d\vec{x}
\label{5.42}
\end{eqnarray}
and
\begin{eqnarray}
S_{lm}(u) &=& - \frac{16\pi i}{(2l+1)!!}\,
\sqrt{ \frac{l+2}{l-1} }\, 
\biggl( \frac{d}{du} \biggr)^l
\int \biggl[ 
1 + \frac{l+4}{2(l+2)(2l+3)}\, (r \partial_u)^2 
\nonumber \\ & & \mbox{} 
- \frac{(l-1)(l+2)}{l}\, \frac{M}{r} + O(v^4) \biggr]
\biggl\{ \biggl[ \mbox{}_{-1} j(u,\vec{x}) - 
\frac{r \partial_u}{l+2}\, \mbox{}_{-1} p(u,\vec{x}) 
\biggr]\, \mbox{}_{-1} \bar{Y}_{lm}(\theta,\phi) 
\nonumber \\ & & \mbox{} 
+ \biggl[ \mbox{}_{1} j(u,\vec{x}) - 
\frac{r \partial_u}{l+2}\, \mbox{}_{1} p(u,\vec{x}) 
\biggr]\, \mbox{}_{1} \bar{Y}_{lm}(\theta,\phi) \biggr\}\,
r^l\, d\vec{x}.
\label{5.43}
\end{eqnarray}
The interpretation of these results is exactly the same as
for the cases (scalar and electromagnetic radiation)
considered previously. Equations (\ref{5.40})--(\ref{5.43}),
with $\beta_l \equiv \beta_l^{\rm grav,current}$ and 
$\mu_l \equiv \beta_l^{\rm grav,mass}$ given by Eqs.~(\ref{5.20})
and (\ref{5.19}), respectively, are equivalent to 
Eqs.~(\ref{1.14})--(\ref{1.19}), once the spin-weighted 
spherical harmonics have been converted into the vectorial
harmonics of Eq.~(\ref{A16}). 

\section{Physical origin of the tail term}

A survey of Secs.~III, IV, and V reveals that for scalar,
electromagnetic, and gravitational radiation, the tail
correction to the radiative multipole moments takes the
universal form [cf.~Eqs.~(\ref{3.20}), (\ref{4.34}), (\ref{4.35}),
(\ref{5.40}), and (\ref{5.41})]
\begin{equation}
{\cal M}_{lm}(u) = M_{lm}(u) + 2M \int_{-\infty}^u \biggl[
\ln \biggl( \frac{u-u'}{4M} \biggr) + c_l + \gamma \biggr]\,
\ddot{M}_{lm}(u')\, du'.
\label{6.1}
\end{equation}
Here, ${\cal M}_{lm}(u)$ stands for ${\cal Z}_{lm}(u)$ in the case 
of scalar radiation, and for ${\cal I}_{lm}(u)$ and ${\cal S}_{lm}(u)$ 
in the case of electromagnetic and gravitational radiation.
Similarly, $M_{lm}(u)$ stands for either $Z_{lm}(u)$, $I_{lm}(u)$, or 
$S_{lm}(u)$. The constant $c_l$ stands for $\beta_l$ 
[cf.~Eq.~(\ref{2.19})], except for the mass multipole 
moments of the gravitational-wave field, for which $c_l$ stands 
for $\mu_l$ [cf.~Eq.~(\ref{5.19})].

The physical interpretation of Eq.~(\ref{6.1}) is clear, and was
first given at the end of Sec.~III C. Equation (\ref{6.1}) shows that 
while the correction terms of order $v^2$ that appear in $M_{lm}(u)$ 
are near-zone corrections that depend on the detailed behavior of
the source, the correction terms of order $v^3$ --- the terms
under the integral sign, or tail terms  --- are due to 
wave-propagation effects, and are independent of the detailed 
behavior of the source. And while the $O(v^2)$ corrections are 
local in time, the $O(v^3)$ corrections introduce a nonlocality 
in the radiative multipole moments. This nonlocality is understood as 
arising from the scattering of the radiation field off the spacetime 
curvature generated by the mass $M$, and as Eq.~(\ref{6.1}) shows, 
the tail term is indeed proportional to $M$.

The mass parameter appears in two places in the Schwarzschild metric,
\begin{equation}
ds^2 = -f dt^2 + f^{-1} dr^2 + r^2 (d\theta^2 + \sin^2\theta\, d\phi^2),
\label{6.2}
\end{equation}
where $f=1-2M/r$. It enters in $g_{tt}$ and in $g_{rr}$, which
are both involved in the calculation of the tail correction. Because 
the mass parameter is the same in both components of the metric, it 
is impossible to tell, on the basis of our previous calculations, 
whether it is $g_{tt}$ that is ``mostly responsible'' for the tail 
effect, or whether it is $g_{rr}$, or whether both components are 
``equally responsible''. In other words, our previous calculations 
cannot tell us how the temporal and spatial curvatures separately 
contribute to the tail effect. This is the question we now wish to 
examine. We shall answer it by artificially introducing an additional
mass parameter in the description of our spacetime. For simplicity, 
we shall restrict attention to the case of scalar radiation. 

We consider a static, spherically symmetric spacetime with
a line element of the most general form,
\begin{equation}
ds^2 = -f dt^2 + g^{-1} dr^2 + r^2 (d\theta^2 + \sin^2\theta\, d\phi^2),
\label{6.3}
\end{equation}
where $f$ and $g$ are two arbitrary functions of $r$ obeying the
following restrictions. First, the spacetime must be asymptotically 
flat, so that the metric functions must behave as
\begin{equation}
f \sim 1 - 2M/r, \qquad
g \sim 1 - 2 \hat{\gamma} M/r,
\label{6.4}
\end{equation}
at large distances ($r \gg M$). Here, $M$ is the gravitational mass 
of the system responsible for the gravitational field, and 
$\hat{\gamma}$ is a parameter that measures the failure of the 
metric to match the Schwarzschild form at large distances; 
$\hat{\gamma} M$ can be thought of as the system's inertial
mass, and the Schwarzschild behavior is recovered by putting 
$\hat{\gamma} = 1$. (This parameter has the same meaning as 
$\gamma$ in the parameterized post-Newtonian formalism \cite{33}. 
We nevertheless use the notation $\hat{\gamma}$ to distinguish 
this quantity from the Euler number $\gamma$.) Second, we assume 
for concreteness that the metric describes a black-hole
spacetime, so that both $f(r)$ and $g(r)$ vanish at a common 
radius $r_0$. Regularity of the spacetime at the event horizon
further demands that the ratio $f/g$ be finite and nonvanishing 
at $r=r_0$. (Our conclusions are insensitive to this second 
set of assumptions.) Apart from these requirements, $f(r)$ and 
$g(r)$ will be left unspecified.

We consider the scalar wave equation, $\Box{} \Phi(\bbox{x})
= -4\pi \rho(\bbox{x})$, in a spacetime with line element 
(\ref{6.3}). After separation of the variables, according
to Eqs.~(\ref{3.2}), (\ref{3.5}), and (\ref{3.6}), the
radial function is found to satisfy 
\begin{equation}
\Biggl\{ \frac{d^2}{dr^{*2}} + \omega^2 - f \Biggl[ 
\frac{l(l+1)}{r^2} + \sqrt{ \frac{g}{f} } 
\frac{ \bigl(\sqrt{fg}\bigr)' }{r} \Biggr] \Biggr\} R_{lm}(\omega;r)
= f T_{lm}(\omega;r),
\label{6.5}
\end{equation}
where $d/dr^* = \sqrt{fg}\, d/dr$, and a prime indicates differentiation
with respect to $r$. This equation is integrated by means of a Green's
function, constructed from two linearly independent solutions to the
homogeneous equation. These are denoted $R^H_{l}(\omega;r)$ and
$R^\infty_{l}(\omega;r)$, and are defined as in 
Eq.~(\ref{3.8}), with $r^* = \int dr/\sqrt{fg}$, 
and with $r=r_0$ replacing $r=2M$. The solution at 
large distances is then given by Eq.~(\ref{3.10}), with
\begin{equation}
\tilde{\cal Z}_{lm}(\omega) = 
\frac{1}{2 i \omega {\cal Q}^{\rm in}_l(\omega)}\,
\int_{r_0}^\infty \sqrt{f/g}\, T_{lm}(\omega;r)\,
R^H_l(\omega;r)\, dr.
\label{6.6}
\end{equation}
Substituting this into Eq.~(\ref{3.2}) yields
\begin{equation}
\Phi_{\rm rad}(t,\vec{x}) = \frac{1}{r} \sum_{lm} {\cal Z}_{lm}(u)\,
Y_{lm}(\theta,\phi),
\label{6.7}
\end{equation}
where ${\cal Z}_{lm}(u)$ is the inverse Fourier transform of
$\tilde{\cal Z}_{lm}(\omega)$, and $u = t-r^*$ is retarded
time. According to Eqs.~(\ref{6.4}), $r^*$ is now given by
\begin{equation}
r^* \sim r + 2\sigma M \ln (r/2M)
\label{6.8}
\end{equation}
at large distances, where
\begin{equation}
\sigma = {\textstyle \frac{1}{2}} \bigl( 1 + \hat{\gamma} \bigr).
\label{6.9}
\end{equation}
Equation (\ref{6.8}) agrees with the Schwarzschild definition
when $\hat{\gamma} = \sigma = 1$. 

We now wish to calculate $\tilde{\cal Z}_{lm}(\omega)$, the
radiative multipole moments, in the slow-motion approximation.
The first step is to integrate the homogeneous version of 
Eq.~(\ref{6.5}) in the low-frequency limit. The calculation
proceeds as in Sec.~II and Appendix C, and uses the approximations
(\ref{6.4}) for $f(r)$ and $g(r)$; these steps will not be duplicated 
here. Defining $\varepsilon = 2M\omega$ and $z=\omega r$, we 
eventually find
\begin{eqnarray}
\frac{R^H_l(z)}{{\cal Q}^{\rm in}_l} &=& 2 (-i)^{l+1} 
e^{i\sigma \varepsilon (\ln 2\varepsilon - \beta_l)} 
\bigl(1+ {\textstyle \frac{\pi}{2}}\sigma \varepsilon\bigr) z 
\Biggl\{ \Bigl[ 1 - \sigma \varepsilon A_l(z) \Bigr] j_l(z) + 
\sigma \varepsilon B_l(z) n_l(z) 
\nonumber \\ & & \mbox{} - 
\varepsilon \frac{(l+1)\hat{\gamma} - \sigma}{2(2l+1)}\, j_{l-1}(z) +
\varepsilon \frac{l\hat{\gamma}+\sigma}{2(2l+1)}\, j_{l+1}(z)
+ O(\varepsilon^2) \Biggr\},
\label{6.10}
\end{eqnarray}
where $A_l(z)$ and $B_l(z)$ are defined by Eqs.~(\ref{2.15})
and (\ref{2.16}), respectively, and
\begin{equation}
\beta_l = \psi(l+1) + \frac{\hat{\gamma}}{2\sigma}.
\label{6.11}
\end{equation}
Evaluating Eq.~(\ref{6.10}) for $z \ll 1$ yields
\begin{eqnarray}
\frac{R^H_l(\omega;r)}{{\cal Q}^{\rm in}_l(\omega)} &=& 
\frac{2}{(2l+1)!!}\, 
e^{2 i \sigma M \omega (\ln 4 M |\omega| - \beta_l)} 
\bigl(1+ \pi \sigma M |\omega|\bigr) (-i\omega r)^{l+1} 
\nonumber \\ & & \mbox{} \times
\Biggl\{
1 - \frac{(\omega r)^2}{2(2l+3)} - 
\Bigl[ (l+1) \hat{\gamma} - \sigma \Bigr]\, \frac{M}{r}
+ O(v^4) \Biggr\}.
\label{6.12}
\end{eqnarray}
The second step is to substitute Eq.~(\ref{6.12}) into
Eq.~(\ref{6.6}), using Eq.~(\ref{6.4}) once more. We
obtain
\begin{eqnarray}
\tilde{\cal Z}_{lm}(\omega) &=& \frac{4\pi}{(2l+1)!!} 
e^{2 i \sigma M \omega (\ln 4 M |\omega| - \beta_l)} 
\bigl(1+ \pi \sigma M |\omega|\bigr) (-i\omega)^l 
\int \biggl[ 1 -
\frac{(\omega r)^2}{2(2l+3)} 
\nonumber \\ & & \mbox{}
- \frac{(2l-1)\hat{\gamma} + 1}{2}\, \frac{M}{r} 
+ O(v^4) \biggr]\, \tilde{\rho}(\omega,\vec{x})\,
r^l\, \bar{Y}_{lm}(\theta,\phi)\, d\vec{x}.
\label{6.13}
\end{eqnarray}
The corresponding expression in the time domain is
\begin{equation}
{\cal Z}_{lm}(u) = Z_{lm}(u) + 2 \sigma M \int_{-\infty}^u 
\biggl[ \ln \biggl( \frac{u-u'}{4M} \biggr)  
+ \beta_l + \gamma \biggr]\,
\ddot{Z}_{lm}(u')\, du',
\label{6.14}
\end{equation}
where 
\begin{equation}
Z_{lm}(u) = \frac{4\pi}{(2l+1)!!} \biggl( \frac{d}{du} \biggr)^l
\int \biggl[ 1 + \frac{(r\partial_u)^2}{2(2l+3)} 
- \frac{(2l-1)\hat{\gamma} + 1}{2}\, \frac{M}{r} 
+ O(v^4) \biggr]\, \rho(u,\vec{x})\, r^l\, \bar{Y}_{lm}(\theta,\phi)\,
d\vec{x}.
\label{6.15}
\end{equation}
These are the radiative multipole moments of a scalar field in
a spacetime with line element (\ref{6.3}), (\ref{6.4}). Equations
(\ref{6.14}) and (\ref{6.15}), with $\beta_l$ given by Eq.~(\ref{6.11}),
are equivalent to Eqs.~(\ref{1.20}) and (\ref{1.21}).

We see from Eqs.~(\ref{6.14}) and (\ref{6.15}) that $\hat{\gamma}$
and $\sigma$ both appear in the near-zone and wave-propagation 
correction terms. In particular, the tail integral is 
now proportional to $\sigma \equiv \frac{1}{2}(1+\hat{\gamma})$. 
This allows us to conclude that in general relativity, temporal 
and spatial curvatures contribute {\it equally} to the 
tail correction. This result is striking, because the same 
conclusion is known to hold in two other situations: the deflection 
and time delay of light by the gravitational field of a massive body. 
Indeed, in a parameterized post-Newtonian calculation \cite{33}, the 
deflection angle and the time delay are both found to be proportional 
to $\frac{1}{2}(1+\gamma)$. (Here, $\gamma$ is the parameter that
measures how much spatial curvature is produced by a unit rest mass;
it is equal to unity in general relativity.) The statement that temporal 
and spatial curvatures contribute equally therefore applies to two very 
different physical situations. While the deflection and time delay of 
light are both high-frequency, geometric-optics phenomena, the tail 
effect is very much a low-frequency, wave-like phenomenon, and the 
discovery of such a similarity in such different situations could not 
have been expected on physical grounds. However, this similarity is not 
entirely surprising on mathematical grounds: The factor of 
$\sigma$ that appears in front of the tail integral is essentially 
the same $\sigma$ that appears in the new definition of $r^*$, 
Eq.~(\ref{6.8}); since radial light rays propagate along curves of 
constant $t-r^*$ or $t+r^*$, it is perhaps not surprising that $\sigma$ 
should also appear in expressions for the deflection angle and the 
time delay.

\section{Spacetime approach}

\subsection{DeWitt-Brehme Green's function}

The mathematical methods employed in the previous sections of this
paper to derive expressions for the radiative multipole moments of 
integer-spin fields were based upon 
a separation of variables approach made possible by the symmetries of the 
Schwarzschild solution. Although the physical interpretation of our 
results, in terms of near-zone and wave-propagation corrections, is 
quite clear, we cannot claim that the physical picture is particularly
well represented by the mathematics involved in bringing the problem to 
a solution. Indeed, the physical meaning of our expressions became clear 
only {\it after} performing the inverse Fourier transform that gave the 
multipole moments in the time domain; by themselves, the frequency-domain 
expressions did not have a very compelling interpretation. 

In this section, we offer an alternative derivation of the radiative 
multipole moments in which the mathematics reflects the physics every 
step of the way. For simplicity we shall again restrict attention to 
the case of scalar radiation.

Following the seminal work by Hadamard \cite{1}, DeWitt and
Brehme \cite{3} considered the scalar wave equation, 
$\Box \Phi(\bbox{x}) = -4\pi \rho(\bbox{x})$, and its
Green's function satisfying $\Box G(\bbox{x},\bbox{x}') =
- \delta(\bbox{x},\bbox{x}')$, in an arbitrary spacetime
with metric $g_{\alpha\beta}$. These equations imply that
the scalar field can be expressed as
\begin{equation}
\Phi(\bbox{x}) = 4\pi \int G(\bbox{x},\bbox{x}')\, \rho(\bbox{x}')\,
d\bbox{x}',
\label{7.1}
\end{equation}
where $d\bbox{x}' = |g(\bbox{x}')|^{1/2}\, d^4 x'$, with 
$g = \det(g_{\alpha\beta})$. Assuming that the field point $\bbox{x}$ 
belongs to the normal convex neighborhood of the source points 
$\bbox{x}'$, DeWitt and Brehme found that the retarded Green's function
takes the form
\begin{equation}
G(\bbox{x},\bbox{x}') = \frac{1}{4\pi}\, \theta(\bbox{x},\bbox{x}')\,
\Bigl[ u(\bbox{x},\bbox{x}')\, \delta(\sigma) - 
v(\bbox{x},\bbox{x}')\, \theta(-\sigma) \Bigr].
\label{7.2}
\end{equation}
Here, $\sigma(\bbox{x},\bbox{x}')$ is the world function first
introduced by Synge \cite{34}, and equal to one-half the squared 
geodesic distance between $\bbox{x}$ and $\bbox{x}'$; $\sigma$ 
is positive if the points are spacelike related, negative if 
the relation is timelike, and zero if $\bbox{x}$ and $\bbox{x}'$ are
joined by a null geodesic. The functions $u(\bbox{x},\bbox{x}')$ 
and $v(\bbox{x},\bbox{x}')$ are nonsingular in the limit $\sigma \to 0$, 
and are obtained by substituting Eq.~(\ref{7.2}) into the differential 
equation for the Green's function. Finally, $\theta(\bbox{x},\bbox{x}')$ 
is a time-ordering function, equal to unity if $\bbox{x}$ is in the 
causal future of $\bbox{x}'$, and zero otherwise.

As can be seen from Eq.~(\ref{7.2}), the retarded Green's function 
splits naturally into a direct part (the first term), which has 
support on, and only on, the past light cone of $\bbox{x}$ (all points
$\bbox{x}'$ such that $\sigma=0$), and a tail part (the second term), 
which has support inside the past light cone (all points
$\bbox{x}'$ such that $\sigma < 0$). This, in turn, 
implies that $\Phi(\bbox{x})$ will also be split into direct and tail 
parts, as was observed in Sec.~III. We therefore see that contrary to
our previous mathematical formulation, Eqs.~(\ref{7.1}) and (\ref{7.2})
reflect the physical picture quite closely. 

In the remainder of this section, we calculate 
$G(\bbox{x},\bbox{x}')$ in a weak-field approximation (relying on 
previous work by DeWitt and DeWitt \cite{35}), and derive an 
expression for the radiative multipole moments of the scalar field. 
Not surprisingly, our answer will agree with what was 
obtained in Sec.~III, Eqs.~(\ref{3.20}) and (\ref{3.21}). Although 
this calculation tells us nothing new in terms of the final answer, 
it is still instructive, because of the fact that the mathematical 
origin of the tail correction is clear from the outset --- it follows 
directly from the tail term in the Green's function.

\subsection{Direct term}

We begin with the calculation of the direct part of the field,
\begin{equation}
\Phi_{\rm direct}(\bbox{x}) = \int \theta(\bbox{x},\bbox{x}')\,
u(\bbox{x},\bbox{x}')\, \delta(\sigma)\, \rho(\bbox{x}')\,
d \bbox{x}'.
\label{7.3}
\end{equation}
This involves the evaluation of $\sigma(\bbox{x},\bbox{x}')$ and 
$u(\bbox{x},\bbox{x}')$. We shall work in the weak-field 
approximation (that is, linearized gravity in harmonic coordinates), 
and express the metric as $g_{\alpha\beta}(\bbox{x}) = 
\eta_{\alpha\beta} + h_{\alpha\beta}(\bbox{x})$, where 
$\eta_{\alpha\beta}$ is the metric of flat spacetime in 
Cartesian coordinates, and
\begin{equation}
h_{\alpha\beta}(\bbox{x}) = \frac{2M}{|\vec{x}|}\, 
\delta_{\alpha\beta}.
\label{7.4}
\end{equation}
Here, $\delta_{\alpha\beta}$ is the Kronecker delta, and
for any three-vector $\vec{s}$, $|\vec{s}|^2 = \vec{s} 
\cdot \vec{s} = \delta_{ab} s^a s^b$. It is assumed that 
the source of the gravitational field is a point mass 
located at the origin of the coordinates. 

The world function is given by \cite{36}
\begin{equation}
\sigma(\bbox{x},\bbox{x}') = \frac{1}{2} \int_{\cal C} 
g_{\alpha \beta}\, \frac{d\xi^\alpha}{d\lambda}\,
\frac{d\xi^\beta}{d\lambda}\, \, d\lambda,
\label{7.5}
\end{equation}
where ${\cal C}$ is the geodesic relating the points $\bbox{x}'$
and $\bbox{x}$, $\xi^\alpha(\lambda)$ the equation of this
geodesic, and $\lambda$ an affine parameter on the 
geodesic, normalized so that $\xi^\alpha(0) = x'^\alpha$ 
and $\xi^{\alpha}(1) = x^\alpha$. Equation (\ref{7.5})
follows immediately from the geometric meaning of the world 
function. 
Because Eq.~(\ref{7.5}) is an action principle for the geodesic 
equation, an error of order $\epsilon$ in the specification of 
$\cal C$ is translated into an error of order $\epsilon^2$ 
in $\sigma(\bbox{x},\bbox{x}')$. Since we wish to evaluate 
$\sigma(\bbox{x},\bbox{x}')$ accurately to {\it first order} 
in the formally small parameter $M$, it is sufficient to 
approximate $\cal C$ by the straight path \cite{36}
\begin{equation}
\xi^\alpha(\lambda) = x^{\prime \alpha} + \lambda \bigl( 
x^\alpha - x^{\prime \alpha} \bigr).
\label{7.6}
\end{equation}
Substituting this into Eq.~(\ref{7.5}) and discarding all
$O(M^2)$ terms, we obtain
\begin{equation}
\sigma(\bbox{x},\bbox{x}') = -\frac{1}{2} \biggl( 1 - 
2M \int_0^1 \frac{d\lambda}{\xi} \biggr) 
\Bigl(t' - t_-\Bigr) \Bigl(t' - t_+ \Bigr),
\label{7.7}
\end{equation}
where
\begin{equation}
t_\pm = t \pm |\vec{x}-\vec{x}'| \pm 2M |\vec{x}-\vec{x}'| 
\int_0^1 \frac{d\lambda}{\xi}
\label{7.8}
\end{equation}
and $\xi \equiv |\vec{\xi}|$. Equation (\ref{7.7}) implies
\begin{equation}
\delta(\sigma) = \frac{1}{|\vec{x}-\vec{x}'|} \biggl[
\delta(t'-t_-) + \delta(t'-t_+) \biggr],
\label{7.9}
\end{equation}
and the second term vanishes when $\delta(\sigma)$ is 
multiplied by $\theta(\bbox{x},\bbox{x}')$. 
To evaluate the integral in Eq.~(\ref{7.8}), we use Eq.~(\ref{7.6}) 
to write
\begin{equation}
\xi = \sqrt{r'^2 + 2 \lambda \vec{x}' \cdot (\vec{x} - \vec{x}') 
+ \lambda^2 |\vec{x} - \vec{x}'|^2},
\label{7.10}
\end{equation}
where $r' = |\vec{x}'|$; we also define $r = |\vec{x}|$ and 
$\vec{n} = \vec{x}/r$. The integration is elementary, and
Eq.~(\ref{7.8}) becomes
\begin{equation}
t_\pm = t \pm |\vec{x}-\vec{x}'| \pm 2M \ln (2r/s'),
\label{7.11}
\end{equation}
where $s' \equiv r' + \vec{n} \cdot \vec{x}'$, and where terms
of order unity have been discarded in the logarithm.  

In the weak-field approximation, $u(\bbox{x},\bbox{x}')$ is 
given by \cite{36}
\begin{equation}
u(\bbox{x},\bbox{x}') = 1 + {\textstyle \frac{1}{2}} 
(x^\alpha - x'^\alpha) (x^\beta - x'^\beta) \int_{\cal C} 
R_{\alpha\beta} \lambda (1-\lambda)\, d\lambda,
\label{7.12}
\end{equation}
where $R_{\alpha\beta}$ is the Ricci tensor. Because $R_{\alpha\beta}$
is already proportional to $M$, the geodesic $\cal C$ can once again
be approximated by the straight path (\ref{7.6}). Now, $R_{\alpha\beta} 
\propto \delta(\vec{x})$ in the point-mass approximation, and there is 
only one path ${\cal C}'$ which gives rise to a nonvanishing integral in 
Eq.~(\ref{7.12}) --- the path for which $\vec{x}'$ and $\vec{x}$ are 
diametrically opposite. Because ${\cal C}'$ forms a set of measure zero 
in the space of all paths connecting source points $\bbox{x}'$ to a given 
field point $\bbox{x}$, the fact that $u(\bbox{x},\bbox{x}') \neq 1$ for
this path has no effect on $\Phi_{\rm direct}(\bbox{x})$. Therefore, we 
can safely set
\begin{equation}
u(\bbox{x},\bbox{x}') = 1
\label{7.13}
\end{equation}
in the following.

Substituting Eqs.~(\ref{7.9}), (\ref{7.11}), and (\ref{7.13})
into Eq.~(\ref{7.3}) yields
\begin{equation}
\Phi_{\rm direct}(t,\vec{x}) = \int 
\frac{\rho(t_-,\vec{x}')}{|\vec{x}-\vec{x}'|}\, d\vec{x}',
\label{7.14}
\end{equation}
where $d\vec{x}' \equiv |g(\bbox{x}')|^{1/2}\, d^3 x = 
(1+2M/r')\, d^3 x$. At large distances, this becomes
\begin{equation}
\Phi^{\rm rad}_{\rm direct}(t,\vec{x}) = \frac{1}{r} \int
\rho \bigl[ u + \vec{n} \cdot \vec{x}' + 2M \ln (s'/2c), \vec{x}' 
\bigr]\, d\vec{x}',
\label{7.15}
\end{equation}
where
\begin{equation}
u = t - r -2M\ln(r/c)
\label{7.16}
\end{equation}
is retarded time, with $c$ an arbitrary constant. This definition 
of retarded time is similar to the Schwarzschild expression, and $c$ 
will eventually be chosen so that the two definitions agree. 

We now invoke the slow-motion approximation and expand $\rho$
in a Taylor series about $u$. (The approximation ensures that 
the series converges.) This gives
\begin{equation}
\Phi^{\rm rad}_{\rm direct}(t,\vec{x}) = \frac{1}{r} \sum_{n=0}^\infty
\frac{1}{n!}\, \int \rho^{(n)}(u,\vec{x}') 
\bigl[ \vec{n} \cdot \vec{x}' + 2M \ln (s'/2c) \bigr]^n\,
d\vec{x}',
\label{7.17}
\end{equation}
where $\rho^{(n)} \equiv \partial^n \rho /\partial u^n$. After 
discarding all terms of second and higher order in $M$, we arrive at
\begin{eqnarray}
\Phi^{\rm rad}_{\rm direct}(t,\vec{x}) &=& \frac{1}{r} \sum_{n=0}^\infty
\frac{1}{n!}\, \int \rho^{(n)}(u,\vec{x}')\,
(\vec{n} \cdot \vec{x}')^n\, d\vec{x}' 
\nonumber \\ & & \mbox{} 
+ \frac{2M}{r} \sum_{n=0}^\infty
\frac{1}{n!}\, \int \rho^{(n+1)}(u,\vec{x}')\, \ln(s'/2c)\, 
(\vec{n} \cdot \vec{x}')^n\, d\vec{x}'.
\label{7.18}
\end{eqnarray}
This is our final expression for the direct part of the
radiative field.

\subsection{Tail term}

The tail part of the scalar field is
\begin{equation}
\Phi_{\rm tail}(\bbox{x}) = - \int \theta(\bbox{x},\bbox{x}')\, 
v(\bbox{x},\bbox{x}')\, \theta(-\sigma)\, \rho(\bbox{x}')\,
d \bbox{x}',
\label{7.19}
\end{equation}
and it is now our task to evaluate this.

An expression for $v(\bbox{x},\bbox{x}')$, accurate to first order 
in $M$ in a weak-field approximation, was derived by DeWitt and 
DeWitt \cite{35}, who find
\begin{eqnarray}
v(\bbox{x},\bbox{x}') &=& -\frac{2 M}{|\vec{x}-\vec{x}'|}\,
\frac{\partial^2}{\partial t' \partial t}\, 
\Biggl[ \theta(r+r'+t'-t) \ln 
\frac{ r+r'+ |\vec{x}-\vec{x}'|}{r+r'-|\vec{x}-\vec{x}'|}
\nonumber \\ & & \mbox{}
+ \theta(t-t'-r'-r) \ln
\frac{ t-t'+ |\vec{x}-\vec{x}'|}{t-t'-|\vec{x}-\vec{x}'|}
\Biggr].
\label{7.20}
\end{eqnarray}
For large $r$, this reduces to
\begin{equation}
v(\bbox{x},\bbox{x}') = -\frac{2M}{r}\, \Biggl[
\frac{\delta(u-u'-s')}{s'} - \frac{\theta(u-u'-s')}{(u-u')^2}
\Biggr],
\label{7.21}
\end{equation}
where $s' = r' + \vec{n} \cdot \vec{x}'$, 
\begin{equation}
u' = t' - r + |\vec{x}-\vec{x}'| \simeq t' - \vec{n} \cdot \vec{x}',
\label{7.22}
\end{equation}
and $u=t-r$ is retarded time. [The true retarded time is given by 
Eq.~(\ref{7.16}) and differs from $t-r$ by a term $2M\ln(r/c)$.
Nevertheless, $u=t-r$ is the appropriate expression to use in the 
calculation of the tail term when working to first order in $M$, 
because $v(\bbox{x},\bbox{x}')$ is already proportional to $M$.]

We now substitute Eq.~(\ref{7.21}) into Eq.~(\ref{7.19}), taking 
into account that $\theta(\bbox{x},\bbox{x}')\, \theta(-\sigma) = 
\theta(t - |\vec{x}-\vec{x}'| -t') = \theta(u-u')$, according
to Eqs.~(\ref{7.7}) and (\ref{7.8}). We obtain
\begin{equation}
\Phi^{\rm rad}_{\rm tail}(t,\vec{x}) = 
\frac{2M}{r}\, \int \int_{-\infty}^u
\biggl[ \frac{\delta(u-u'-s')}{s'} - 
\frac{\theta(u-u'-s')}{(u-u')^2} \biggr]\,
\rho(u' + \vec{n}\cdot \vec{x}',\vec{x}')\, du'\, d\vec{x}',
\label{7.23}
\end{equation}
where $d\vec{x}' = |g(\bbox{x}')|^{1/2}\, d^3 x' = 
[1+O(M)]\, d^3 x'$. After two partial integrations and 
a few lines of algebra, this becomes
\begin{eqnarray}
\Phi^{\rm rad}_{\rm tail}(t,\vec{x}) &=& \frac{2M}{r}\, \int \biggl[
- \dot{\rho}(u-r',\vec{x}')\, \ln s' - 
\int_{-\vec{n} \cdot \vec{x}'}^{r'} \ddot{\rho}(u-\zeta,\vec{x}')\,
\ln (\zeta+\vec{n} \cdot \vec{x}')\, d\zeta 
\nonumber \\ & & \mbox{} 
+ \int_{-\infty}^u \ddot{\rho}(u' + \vec{n} \cdot \vec{x}',\vec{x}')\,
\ln (u-u')\, du' \biggr]\, d\vec{x},
\label{7.24}
\end{eqnarray}
where a dot indicates differentiation with respect to either $u$
or $u'$. 

The $\zeta$-integral to the right-hand side of Eq.~(\ref{7.24})
can be evaluated explicitly if $\ddot{\rho}(u-\zeta,\vec{x}')$
is expanded in a Taylor series about $\zeta=0$. (Again, the 
slow-motion approximation ensures that this series converges.)
This results in an infinite sum of terms involving the integrals 
$\int \zeta^n \ln (\zeta+\vec{n} \cdot \vec{x}')\, d\zeta$,
which can be expressed in closed form (Ref.~\cite{37}, p.~205). 
After a rather long but straightforward calculation, we find 
that this integral is equal to
\begin{eqnarray}
\dot{\rho}(u-r',\vec{x}')\, \ln s' + \sum_{n=0}^\infty 
\frac{1}{n!}\, \Biggl\{ & & 
\Bigl[ -\ln s' + \psi(n+1) + \gamma \Bigr]\,
\rho^{(n+1)}(u,\vec{x}') 
\nonumber \\ & & \mbox{}
- \sum_{p=n+1}^\infty 
\frac{(-1)^{p-n} n!}{p!(p-n)}\, \rho^{(p+1)}(u,\vec{x}')\, r'^{p-n}
\Biggr\}\, (\vec{n} \cdot \vec{x}')^n.
\label{7.25}
\end{eqnarray}
Notice that the first term cancels out the first term to the
right-hand side of Eq.~(\ref{7.24}). The rest of Eq.~(\ref{7.25}) 
is simplified by invoking the slow-motion approximation. Because 
it involves an additional (retarded) time derivative, the first 
term in the sum over $p$ is smaller than $\rho^{(n+1)}$ by a factor 
of order $v$, and the remaining terms are smaller still. Now, 
$M \rho^{(n+1)}$ is already a factor of order $v^3$ smaller than 
$\rho^{(n)}$, which appears in the direct part of the radiative 
field. This means that in Eq.~(\ref{7.25}), the sum over $p$ is 
$O(v^4)$, and therefore, it will be neglected. 

After substituting Eq.~(\ref{7.25}) into Eq.~(\ref{7.24}), and
expanding the third term to the right-hand side of this 
equation in a Taylor series about $\vec{n} \cdot \vec{x}' = 0$,
we arrive at
\begin{eqnarray}
\Phi^{\rm rad}_{\rm tail}(t,\vec{x}) &=& \frac{2M}{r} \sum_{n=0}^\infty
\frac{1}{n!}\, \int \biggl\{ 
\Bigl[ -\ln s' + \psi(n+1) + \gamma \Bigr]\, \rho^{(n+1)}(u,\vec{x}')
\nonumber \\ & & \mbox{}
+ \int_{-\infty}^u \rho^{(n+2)}(u',\vec{x}')\, \ln(u-u')\, du'
\biggr\}\, (\vec{n} \cdot \vec{x}')^n\, d\vec{x}'.
\label{7.26}
\end{eqnarray}
This is our final expression for the tail part of the radiative field.

\subsection{Total radiative field}

The total radiative field is obtained by adding the direct and 
tail terms. Combining Eqs.~(\ref{7.18}) and (\ref{7.26}), we
obtain
\begin{eqnarray}
\Phi_{\rm rad}(t,\vec{x}) &=& 
\frac{1}{r} \sum_{n=0}^\infty \frac{1}{n!}\,
\int \biggl\{ \rho^{(n)}(u,\vec{x}') + 2M \int_{-\infty}^u
\biggl[ \ln \biggl(\frac{u-u'}{2c}\biggr) 
\nonumber \\ & & \mbox{}
+ \psi(n+1) + 
\gamma \biggr]\, \rho^{(n+2)}(u',\vec{x}')\, du' \biggr\} 
(\vec{n} \cdot \vec{x}')^n\, d\vec{x}'.
\label{7.27}
\end{eqnarray}
We recall that $d\vec{x}' = (1+2M/r')d^3 x'$, $u=t-r-2M\ln(r/c)$,
and that Eq.~(\ref{7.27}) has been derived on the basis of a 
weak-field, slow-motion approximation; it is valid to first 
order in $M$, and neglects terms of order $v^4$. 

The constant $c$ appearing in Eq.~(\ref{7.27}) is the same one which 
enters in the definition of the retarded time $u$, Eq.~(\ref{7.16}).
The radiative field does not actually depend on the numerical value 
of this constant. To see this, let $c \to \lambda c$, where $\lambda$ 
is a scaling constant. Then Eq.~(\ref{7.16}) implies $u \to u + 
2M\ln \lambda$, and we have $\rho^{(n)}(u,\vec{x}') \to 
\rho^{(n)}(u,\vec{x}') + 2M\ln \lambda \rho^{(n+1)}(u,\vec{x}') + O(M^2)$.
Substituting these relations into Eq.~(\ref{7.27}) and discarding
all terms of order $M^2$ shows that indeed, $\Phi_{\rm rad}(\bbox{x})$ 
is invariant under this transformation. 

Our current expression for the radiative field has a mathematical 
structure similar to that of Eqs.~(\ref{3.12}), (\ref{3.20}), and 
(\ref{3.21}), but there appear to be some differences. We now 
show that these are only apparent, and that in fact, Eq.~(\ref{7.27}) 
is entirely equivalent to the results of Sec.~III.

We first re-introduce the spherical harmonics, with
the relation \cite{38}
\begin{equation}
(\vec{n}\cdot \vec{x}')^n = 4\pi\, n!\, r'^n\, 
\mathop{{\sum}'}_{l=0}^n
\sum_{m=-l}^l \frac{2^l \bigl( \frac{n+l}{2} \bigr)!}{(n+l+1)!
\bigl( \frac{n-l}{2} \bigr)! }\, \bar{Y}_{lm}(\theta',\phi')\,
Y_{lm}(\theta,\phi),
\label{7.28}
\end{equation}
where the sum over $l$ includes even values only if $n$ is even,
and odd values only if $n$ is odd. The angles $\theta'$ and $\phi'$
are the polar angles of the source point $\vec{x}'$, and $\theta$ 
and $\phi$ belong to the field point $\vec{x}$. Substituting this 
into Eq.~(\ref{7.27}) and reordering the sums, we obtain
\begin{eqnarray}
\Phi(t,\vec{x}) &=& \frac{4\pi}{r} \sum_{l=0}^\infty \sum_{m=-l}^l
\bar{Y}_{lm}(\theta',\phi')\, Y_{lm}(\theta,\phi)\,
\mathop{{\sum}'}_{n=l}^\infty 
\frac{2^l \bigl( \frac{n+l}{2} \bigr)!}{(n+l+1)!
\bigl( \frac{n-l}{2} \bigr)! }\, \int \biggl\{
\rho^{(n)}(u,\vec{x}') 
\nonumber \\ & & \mbox{}
+ 2M \int_{-\infty}^u
\biggl[ \biggl(\ln \frac{u-u'}{2c} \biggr) + \psi(n+1) + 
\gamma \biggr]\, \rho^{(n+2)}(u',\vec{x}')\, du' \biggr\} 
r'^n\, d\vec{x}'.
\label{7.29}
\end{eqnarray}
The slow-motion approximation now demands that we keep only
the first two terms ($n=l$ and $n=l+2$) in the sum over $n$.  
After some algebra, we arrive at 
\begin{equation}
\Phi_{\rm rad}(t,\vec{x}) = \frac{1}{r} \sum_{lm} {\cal Z}_{lm}(u)\,
Y_{lm}(\theta,\phi),
\label{7.30}
\end{equation}
which is the same as Eq.~(\ref{3.12}). Here,
\begin{equation}
{\cal Z}_{lm}(u) = Z_{lm}(u) + 2M \int_{-\infty}^u \biggl[
\ln \biggl( \frac{u-u'}{2c} \biggr) + \psi(l+1) + \gamma \biggr]\,
\ddot{Z}_{lm}(u')\, du',
\label{7.31}
\end{equation}
and 
\begin{equation}
Z_{lm}(u) = \frac{4\pi}{(2l+1)!!} \biggl( \frac{d}{du} \biggr)^l
\int \biggl[ 1 + \frac{(r\partial_u)^2}{2(2l+3)} + O(v^4) 
\biggr]\, \rho(u,\vec{x})\, r^l\, \bar{Y}_{lm}(\theta,\phi)\,
d\vec{x}.
\label{7.32}
\end{equation}
This is almost, but not quite, the same as Eqs.~(\ref{3.20}) and
(\ref{3.21}). 

To properly compare our results with those of Sec.~III, we
must account for the different choices of coordinate systems. The
coordinates used in this section, and those for which Eq.~(\ref{7.4}) 
holds, are the harmonic coordinates $\{t,x,y,z\}$. From these we
have constructed the spherical coordinates $\{t,r,\theta,\phi\}$
in the usual way, and in this coordinate system, $d\vec{x} =
(1+2M/r) r^2 dr\, d\Omega$, where $d\Omega = d\cos\theta\, d\phi$.
These coordinates are distinct from the Schwarzschild coordinates 
used in Sec.~III, which we now denote $\{\bar{t},\bar{r},\bar{\theta},
\bar{\phi}\}$. The transformation between the two coordinate systems 
is \cite{39}
\begin{equation}
\bar{t} = t, \qquad \bar{r} = r+M, \qquad 
\bar{\theta} = \theta, \qquad \bar{\phi} = \phi.
\label{7.33}
\end{equation}
We therefore have $d\vec{x} = \bar{r}^2 d\bar{r}\, d\bar{\Omega} 
\equiv d\vec{\bar{x}}$, which is the volume element of Sec.~III. 
We also have $r^l = \bar{r}^l (1 - lM/\bar{r})$, and substituting 
this into Eq.~(\ref{7.31}) yields
\begin{equation}
Z_{lm}(u) = \frac{4\pi}{(2l+1)!!} \biggl( \frac{d}{du} \biggr)^l
\int \biggl[ 1 + \frac{(\bar{r}\partial_u)^2}{2(2l+3)} - 
l \frac{M}{\bar{r}} + O(v^4) \biggr]\, 
\rho(u,\vec{\bar{x}})\, \bar{r}^l\, 
\bar{Y}_{lm}(\bar{\theta},\bar{\phi})\, d\vec{\bar{x}}.
\label{7.34}
\end{equation}
This is the same as Eq.~(\ref{3.21}).

Finally, a specific choice for $c$ can be made by demanding that
$u=t-r-2M\ln(r/c)$ be equal to $\bar{u} = \bar{t} - \bar{r}
-2M \ln(\bar{r}/2M)$, which is the retarded time encountered in 
Sec.~III. (We have approximated $\bar{r}/2M - 1$ by $\bar{r}/2M$ 
in the logarithm.) A short calculation gives
\begin{equation}
c = 2M e^{-1/2},
\label{7.35}
\end{equation}
and with this choice, Eq.~(\ref{7.31}) becomes
\begin{equation}
{\cal Z}_{lm}(u) = Z_{lm}(u) + 2M \int_{-\infty}^u \biggl[
\ln \biggl( \frac{u-u'}{4M} \biggr) + \psi(l+1) + 
\frac{1}{2} + \gamma \biggr]\, \ddot{Z}_{lm}(u')\, du'.
\label{7.36}
\end{equation}
This is the same as Eq.~(\ref{3.20}). We therefore have 
precise agreement between the results of this section and
those of Sec.~III.

\section*{Acknowledgments}

Conversations with Ray McLenaghan and Bernie Nickel were greatly 
appreciated. This work was supported by the Natural Sciences and 
Engineering Research Council of Canada. Steve Leonard's work was 
also supported by an Ontario Graduate Scholarship.

\appendix
\section{Tensor fields on a two-sphere}

We gather, for the benefit of the reader, several results pertaining 
to the ``edth'' differential operators and the associated 
spin-weighted spherical harmonics. The discussion follows 
closely Ref.~\cite{40}, but it is essentially self-contained. 

We consider $S^2$, a spherical two-dimensional space with metric
\begin{equation}
ds^2 = g_{ab} d\theta^a d\theta^b = r^2(d\theta^2 + \sin^2\theta\, d\phi^2),
\label{A1}
\end{equation}
on which fields of various tensorial types are defined. For simplicity,
all geometric objects considered in this Appendix will be confined to 
this space. However, extension of the following considerations to 
four-dimensional, spherically symmetric spacetimes is immediate.

It is useful to introduce a set of basis vectors, $m^a$ and
$\bar{m}^a$ (a bar denotes complex conjugation), which
satisfy the relations
\begin{equation}
m_a m^a = \bar{m}_a \bar{m}^a = 0,
\qquad
m_a \bar{m}^a = 1.
\label{A2}
\end{equation}
A particular choice is
\begin{equation}
m_a = \frac{r}{\sqrt{2}}\, (1, i\sin\theta), \qquad
\bar{m}_a = \frac{r}{\sqrt{2}}\, (1,-i\sin\theta).
\label{A3}
\end{equation}
This choice is not unique, because the relations (\ref{A2}) 
are preserved under the transformation
\begin{equation}
m^a \to e^{i\psi} m^a, \qquad
\bar{m}^a \to e^{-i\psi} \bar{m}^a,
\label{A4}
\end{equation}
where $\psi$ is any constant. 

We may use the basis to decompose tensor fields. For example,
a vector field $V^a$ may be expressed as 
\begin{equation}
V_{a} = \mbox{}_{-1} V\, m_a + \mbox{}_{1} V\, \bar{m}_a,
\label{A5}
\end{equation}
where
\begin{equation}
\mbox{}_{1} V = V^a m_a, \qquad
\mbox{}_{-1} V = V^a \bar{m}_a.
\label{A6}
\end{equation}
Similarly, a symmetric tensor field $T^{ab}$ is decomposed as
\begin{equation}
T_{ab} = \mbox{}_{-2} T\, m_a m_b + 2\, \mbox{}_{0} T\, m_{(a} \bar{m}_{b)}
+ \mbox{}_{2} T\, \bar{m}_a \bar{m}_b,
\label{A7}
\end{equation}
where
\begin{equation}
\mbox{}_{2}T = T^{ab} m_a m_b, \qquad
\mbox{}_{0}T = T^{ab} m_a \bar{m}_b, \qquad
\mbox{}_{-2}T = T^{ab} \bar{m}_a \bar{m}_b.
\label{A8}
\end{equation}

The {\it spin-weight} of a field is determined by how the field 
transforms under (\ref{A4}). By definition, a field has spin-weight
$s$, and is denoted $\mbox{}_{s} \eta$, if
\begin{equation}
\mbox{}_s \eta \to e^{i s \psi} \mbox{}_s \eta
\label{A9}
\end{equation}
under the transformation. For example, $\mbox{}_{-1} V$ has
spin-weight $s=-1$, while $\mbox{}_2 T$ has spin-weight $s=2$.

The covariant derivatives (with respect to $g_{ab}$) of the base 
vectors are given by
\begin{equation}
m_{a;b} = -\frac{1}{\sqrt{2}r} \cot\theta\, m_a (m_b - \bar{m}_b)
\label{A10}
\end{equation}
and its complex conjugate. 
It follows that the covariant derivatives of arbitrary tensor
fields can be conveniently expressed in terms of the ``edth'' 
differential operators, $\hat{\partial}$ and $\check{\partial}$,
which are defined by
\begin{equation}
\hat{\partial} = -\biggl( \frac{\partial}{\partial\theta}
+ i \csc\theta \frac{\partial}{\partial\phi} - s \cot\theta \biggr),
\qquad
\check{\partial} = -\biggl( \frac{\partial}{\partial\theta}
- i \csc\theta \frac{\partial}{\partial\phi} + s \cot\theta \biggr).
\label{A11}
\end{equation}
It should be noted that these operators depend on $s$, the spin-weight
of the object on which they act. (The original notation \cite{40}
for these operators was $\partial^{\!\!\!-}\!$ and 
$\bar{\partial}^{\!\!\!-}\!$, respectively.) For example, 
\begin{equation}
V_{a;b} = -\frac{1}{\sqrt{2}r} \Bigl[ 
\bigl( \check{\partial} \mbox{}_{-1} V \bigr) m_a m_b + 
\bigl( \hat{\partial} \mbox{}_{-1} V \bigr) m_a \bar{m}_b + 
\bigl( \check{\partial} \mbox{}_{1} V \bigr) \bar{m}_a m_b + 
\bigl( \hat{\partial} \mbox{}_{1} V \bigr) \bar{m}_a \bar{m}_b \Bigr].
\label{A12}
\end{equation}
From this relation it is clear that $\hat{\partial}$ raises
the spin-weight by one unit, while $\check{\partial}$ lowers it
by one unit. For example, $\hat{\partial} \mbox{}_1 V =
- \sqrt{2} r V_{a;b}m^a m^b$ has spin-weight $s=2$.

The ``edth'' operators can be manipulated efficiently when working 
under an integral sign. Given two smooth, complex functions 
$\mbox{}_{s-1}f(\theta,\phi)$ and $\mbox{}_s g(\theta,\phi)$, the 
following identities are easily established by straightforward partial 
integration:
\begin{equation}
\int (\hat{\partial} \mbox{}_{s-1}f) \mbox{}_s \bar{g}\, d\Omega = -
\int \mbox{}_{s-1}f (\overline{ \check{\partial} \mbox{}_{s}g })\, 
d\Omega, \qquad
\int (\check{\partial} \mbox{}_{s}g) \mbox{}_{s-1}\bar{f} \, d\Omega = -
\int \mbox{}_{s}g (\overline{ \hat{\partial} \mbox{}_{s-1}f })\, d\Omega,
\label{A13}
\end{equation}
where $d\Omega = d\cos\theta\, d\phi$. 

The ``edth'' operators can be used to generate sets of spin-weighted
spherical-harmonic functions, denoted $\mbox{}_s Y_{lm}(\theta,\phi)$.
Each set (corresponding to a fixed value of $s$) is complete, and
members of a given set obey the usual orthonormality relations. The 
defining relations are $\mbox{}_{0} Y_{lm} \equiv Y_{lm}$ 
(the usual spherical harmonics), and
\begin{equation}
\hat{\partial} \mbox{}_s Y_{lm} = \sqrt{(l-s)(l+s+1)}\, 
\mbox{}_{s+1} Y_{lm}, \qquad
\check{\partial} \mbox{}_s Y_{lm} = -\sqrt{(l+s)(l-s+1)}\, 
\mbox{}_{s-1} Y_{lm}.
\label{A14}
\end{equation}
The spin-weighted spherical harmonics also satisfy the relations
\begin{equation}
\mbox{}_{-s} \bar{Y}_{l,-m} = (-1)^{s+m} \mbox{}_s Y_{lm}.
\label{A15}
\end{equation}

The spin-weighted spherical harmonics can be combined with 
basis vectors to form tensorial spherical harmonics \cite{25}. 
For example,
\begin{equation}
Y^{E,lm}_a = \frac{1}{\sqrt{2}} \bigl( \mbox{}_{-1} Y_{lm}\, m_a 
- \mbox{}_1 Y_{lm}\, \bar{m}_a \bigr), \qquad
Y^{B,lm}_a = - \frac{i}{\sqrt{2}} \bigl( \mbox{}_{-1} Y_{lm}\, m_a 
+ \mbox{}_1 Y_{lm}\, \bar{m}_a \bigr),
\label{A16}
\end{equation}
are vectorial spherical harmonics. The superscript $E$
indicates that under a parity transformation, $Y^{E,lm}_a$ has
electric-type parity: $Y^{E,lm}_a \to (-1)^l Y^{E,lm}_a$; 
the superscript $B$ indicates a magnetic-type parity: 
$Y^{B,lm}_a \to (-1)^{l+1} Y^{B,lm}_a$. Similarly, 
\begin{equation}
T^{E2,lm}_{ab} = \frac{1}{\sqrt{2}} \bigl( 
\mbox{}_{-2} Y_{lm}\, m_a m_b + 
\mbox{}_2 Y_{lm}\, \bar{m}_a \bar{m}_b \bigr), \qquad
T^{B2,lm}_{ab} = -\frac{i}{\sqrt{2}} \bigl( 
\mbox{}_{-2} Y_{lm}\, m_a m_b - 
\mbox{}_2 Y_{lm}\, \bar{m}_a \bar{m}_b \bigr),
\label{A17}
\end{equation}
are tensorial spherical harmonics.
 
\section{Evaluation of two functions}

We evaluate, in the limit $z \to 0$,  the functions $A_l(z)$ and
$B_l(z)$ defined by Eqs.~(\ref{2.15}) and (\ref{2.16}).

To evaluate $A_l(z)$ is easy. By using the expansions 
${\rm Si}(2z) = 2z + O(z^3)$ and $z^2 n_p j_p = 
- z/(2p+1) + O(z^3)$, we quickly arrive at
\begin{equation}
A_l(z) = z - z \sum_{p=1}^{l-1} 
\biggl(\frac{1}{p} - \frac{1}{p+1} \biggr) + O(z^3).
\label{B1}
\end{equation}
The sum evaluates to $1-1/l$, and $A_l(z)$ reduces to the
result quoted in the text --- Eq.~(\ref{2.17}).

To evaluate $B_l(z)$ requires more work.
We begin by recalling the series expansions for the cosine 
integral (Ref.~\cite{29}, p.~232),
\begin{equation}
{\rm Ci}(2z) = \gamma + \ln(2z) + \sum_{n=1}^\infty 
\frac{(-1)^n}{2n(2n)!}\, (2z)^{2n},
\label{B2}
\end{equation}
the squared sine,
\begin{equation}
\sin^2 z = - \sum_{n=1}^\infty \frac{(-1)^n}{2(2n)!}\,
(2z)^{2n},
\label{B3}
\end{equation}
and the squared spherical Bessel functions of the first kind 
(Ref.~\cite{37}, p.~960),
\begin{equation}
z^2 {j_p}^2(z) = \pi \sum_{k=0}^\infty 
\frac{(-1)^k \Gamma(2p+2+2k)}{k! \Gamma(2p+2+k) 
\Gamma^2(p + \frac{3}{2} + k)}\, 
\Bigl( \frac{z}{2} \Bigr)^{2p + 2 + 2k}.
\label{B4}
\end{equation}
Substituting these into Eq.~(\ref{2.16}) gives the series
\begin{equation}
B_l(z) = \sum_{n=1}^\infty b_n z^{2n},
\label{B5}
\end{equation}
where, after some rearranging,
\begin{equation}
b_n = \frac{(-1)^{n+1} 2^{2n} (n-1)}{2n (2n)!} + 
\frac{\pi \Gamma(2n)}{2^{2n} \Gamma^2(n+\frac{1}{2})}\,
\sum_{p=1}^P \biggl(\frac{1}{p} + \frac{1}{p+1} \biggr)
\frac{(-1)^{n-p-1}}{(n-p-1)! \Gamma(n+p+1)},
\label{B6}
\end{equation}
with $P=\min(n-1,l-1)$. Additional manipulations bring $b_n$
to the form
\begin{equation}
b_n = \frac{ (-1)^{n+1} 2^{2n} (n-1) }{ 2n (2n)! }\,
\Biggl[ 1 + \frac{n!^2}{n-1} 
\sum_{p=1}^P  
\biggl(\frac{1}{p} + \frac{1}{p+1} \biggr) 
\frac{(-1)^p}{(n-p-1)! (n+p)!} \Biggr].
\label{B7}
\end{equation}
The sum evaluates to
\begin{equation}
- \frac{n-1}{n!^2} + \frac{(-1)^P (n-P-1)}{n (P+1) (n+P)! (n-P-1)!},
\label{B8}
\end{equation}
which implies that $b_n=0$ for $n \leq l$ (because $P=n-1$),
while
\begin{equation}
b_n = \frac{(-1)^{n+l} 2^{2n-1} (n-1)!^2}{l (2n)! (n-l-1)! 
(n+l-1)!}
\label{B9}
\end{equation}
for $n \geq l+1$. It is then easy to show that Eq.~(\ref{B5}) 
reduces to the result quoted in the text --- Eq.~(\ref{2.17}).

\section{Regge-Wheeler function in the asymptotic limit}

We wish to evaluate the Regge-Wheeler function $X^H_l(z)$, as
given by Eqs.~(\ref{2.9}) and (\ref{2.14}) (with $a=0$ and 
$b=-\gamma$), in the asymptotic limit $z \to \infty$. By comparing
with the low-frequency limit of Eq.~(\ref{2.5}), we will then
compute ${\cal A}^{\rm in}_l$ in the normalization provided by 
Eq.~(\ref{2.9}). 

To express Eqs.~(\ref{2.9}) and (\ref{2.14}) in the limit $z\to \infty$,
we use such asymptotic relations as ${\rm Si}(2z) \sim \pi/2$, 
${\rm Ci}(2z) \sim 0$, $z^3 (n_l j_p - j_l n_p) j_p \sim \frac{1}{2}
[ 1 - (-1)^{l-p} ] z n_l$, $j_{l-1} \sim - n_l$, and $j_{l+1} \sim
n_l$. After some algebra, we obtain
\begin{equation}
X^H_l \sim \bigl(1 - \varepsilon {\textstyle \frac{\pi}{2}} \bigr) 
z j_l - \varepsilon \bigl[ \ln(2z) - \beta_l \bigr] z n_l + 
O(\varepsilon^2),
\label{C1}
\end{equation}
where $\beta_l$ is given by Eq.~(\ref{2.19}).

We must now compare this result with the low-frequency limit of
Eq.~(\ref{2.5}), which we rewrite as
\begin{equation}
X^H_l \sim {\cal A}^{\rm in}_l e^{-i z^*} + 
{\cal A}^{\rm out}_l e^{i z^*},
\label{C2}
\end{equation}
where $z^* = z + \varepsilon \ln (z/\varepsilon -1)$. 
Expanding the phase factors in powers of $\varepsilon$, and
using the asymptotic relations $e^{\pm i z} \sim (\pm i)^{l+1}
( z j_l \pm i z n_l )$, yields
\begin{equation}
X^H_l \sim (1+\varepsilon {\cal A}^+_l) z j_l +
\varepsilon ({\cal A}^-_l - \ln z) z n_l + O(\varepsilon^2),
\label{C3}
\end{equation}
where
\begin{eqnarray}
1 + \varepsilon {\cal A}^+_l &=& (-i)^{l+1} {\cal A}^{\rm in}_l
e^{i \varepsilon \ln \varepsilon} +  
(i)^{l+1} {\cal A}^{\rm out}_l e^{-i \varepsilon \ln \varepsilon},
\nonumber \\
& & \label{C4} \\
i \varepsilon {\cal A}^-_l &=& (-i)^{l+1} {\cal A}^{\rm in}_l
e^{i \varepsilon \ln \varepsilon} -  
(i)^{l+1} {\cal A}^{\rm out}_l e^{-i \varepsilon \ln \varepsilon}.
\nonumber
\end{eqnarray}
Finally, comparing Eqs.~(\ref{C1}) and (\ref{C3}), and using the 
relations (\ref{C4}), we arrive at
\begin{equation}
{\cal A}^{\rm in}_l = {\textstyle \frac{1}{2}} (i)^{l+1} 
e^{-i \varepsilon (\ln 2 \varepsilon - \beta_l)} \Bigl[
1 - {\textstyle \frac{\pi}{2}} \varepsilon + O(\varepsilon^2) \Bigr].
\label{C5}
\end{equation}
From this and Eqs.~(\ref{2.9}), (\ref{2.14}), we obtain 
Eq.~(\ref{2.18}).

\section{Inverse Fourier transform of tail corrections}

We wish to take the inverse Fourier transform of the function
\begin{equation}
\tilde{\cal F}(\omega) = e^{2iM\omega (\ln 4M|\omega| - c)} 
\bigl( 1 + \pi M |\omega| \bigr)\, \tilde{F}(\omega),
\label{D1}
\end{equation}
where $c$ is a constant and $\tilde{F}(\omega)$ an arbitrary,
square-integrable function. In other words, we wish to compute 
the function ${\cal F}(u)$ given by
\begin{equation}
{\cal F}(u) = \int \tilde{\cal F}(\omega) e^{-i\omega u}\, d\omega.
\label{D2}
\end{equation}
We shall do so in the spirit of the slow-motion approximation, by 
formally treating $M$ as a small parameter. We follow closely 
the derivation found in Appendix A of Ref.~\cite{16}. 

We first expand the exponential factor in Eq.~(\ref{D1}) to linear 
order in $M$, and combine the result with the $(1+\pi M |\omega|)$ 
factor. We then substitute the identity $i\omega \ln|\omega| + 
\pi |\omega|/2 = i\omega \ln(-i\omega)$. After a few lines of algebra,
we obtain
\begin{equation}
{\cal F}(u) = F(u) - 2M(\ln 4M - c) \dot{F}(u) + 
2M \int \tilde{F}(\omega)\, i\omega \ln(-i\omega) 
e^{-i\omega u}\, d\omega,
\label{D3}
\end{equation}
where $F(u)$ is the inverse Fourier transform of $\tilde{F}(\omega)$,
and a dot indicates differentiation with respect to $u$.

To evaluate the integral, we write $\ln(-i\omega)$ in a
different form by using the identity (Ref.~\cite{37}, p.~573)
\begin{equation}
\ln \mu = -\gamma -\mu \int_0^\infty e^{-\mu x} \ln x\, dx,
\label{D4}
\end{equation}
with $\mu = -i\omega$; $\gamma \simeq 0.57721$ is Euler's number. 
Strictly speaking, this identity is valid only if the real part 
of $\mu$ is positive. This problem can be circumvented by introducing 
a regulator $\epsilon > 0$, and setting $\mu = -i\omega + \epsilon$. 
The limit $\epsilon \to 0$ can be taken after integrating over $\omega$, 
which yields
\begin{equation}
{\cal F}(u) = F(u) - 2M(\ln 4M - c - \gamma)\, \dot{F}(u) + 
2M \int_0^\infty \ln x\, \ddot{F}(u-x)\, dx.
\label{D5}
\end{equation}
We write this in its final form as
\begin{equation}
{\cal F}(u) = F(u) + 2M \int_{-\infty}^u \biggl[
\ln \biggl( \frac{u-u'}{4M} \biggr) + c + \gamma \biggr]\,
\ddot{F}(u')\, du'.
\label{D6}
\end{equation}
This is the desired result.

\section{Chandrasekhar transformation for the electromagnetic field}

We derive the relation between $R^H_l(\omega;r)$, the solution to
the homogeneous version of Eq.~(\ref{4.2}) with boundary
conditions (\ref{4.9}), and $X^H_l(\omega;r)$, the solution to 
the $s=1$ version of Eq.~(\ref{2.1}) with boundary 
conditions (\ref{2.4})--(\ref{2.5}). For convenience, in this 
Appendix we set to unity the arbitrary constant appearing in 
Eq.~(\ref{2.4}). 

Direct substitution shows that if $X_l(\omega;r)$ satisfies the 
generalized Regge-Wheeler equation (with $s=1$), then
$R_l(\omega;r) = r {\cal L} X_l(\omega;r)$ satisfies the 
homogeneous Teukolsky equation. Here, ${\cal L} = 
fd/dr + i\omega$. The desired relation must 
therefore have the form
\begin{equation}
R^H_l(\omega;r) = \chi\, r {\cal L} X^H_l(\omega;r),
\label{E1}
\end{equation}
where the constant $\chi$ must be chosen so that the
normalization of $R^H_l(\omega;r)$ agrees with Eq.~(\ref{4.9}).

To find $\chi$ and to relate ${\cal Q}^{\rm in}_l(\omega)$ to
${\cal A}^{\rm in}_l(\omega)$, we need expressions for 
$X^H_l(\omega;r)$ that are more accurate than Eqs.~(\ref{2.4})
and (\ref{2.5}). By solving the generalized Regge-Wheeler
equation, we find that  
\begin{equation}
X^H_l(\omega;r) = \biggl[ 1 + \frac{l(l+1)}{1-4iM\omega}\, f
+ O(f^2) \biggr] e^{-i\omega r^*}
\label{E2}
\end{equation}
near $r=2M$, while
\begin{equation}
X^H_l(\omega;r) = {\cal A}^{\rm in}_l(\omega) \biggl\{
1 + \frac{l(l+1)}{2i\omega r} + O\bigl[ (\omega r)^{-2} \bigr] \biggr\}
e^{-i\omega r^*} + \cdots
\label{E3}
\end{equation}
near $r=\infty$, where the dots designate terms proportional to 
$e^{i\omega r^*}$. We are now in a position to verify that near 
$r=2M$, $r {\cal L} e^{-i\omega r^*} \sim l(l+1) f 
e^{-i\omega r^*} / (1-4iM\omega)$, and that near
$r=\infty$, $r {\cal L} e^{-i\omega r^*} = -\frac{1}{2} l(l+1) 
(i\omega r)^{-1} e^{-i\omega r^*}$.

Combining these results with Eqs.~(\ref{2.4}), (\ref{2.5}),
(\ref{4.9}), and (\ref{E1}), we find
\begin{equation}
\chi = \frac{1 - 4iM\omega}{l(l+1)}
\label{E4}
\end{equation}
and
\begin{equation}
{\cal Q}^{\rm in}_l(\omega) = - \frac{1}{2}\, (1 - 4iM\omega) \,
{\cal A}^{\rm in}_l(\omega).
\label{E5}
\end{equation}
Equation (\ref{4.15}) follows immediately.

\end{document}